\documentclass[aps,prl,reprint,amsmath,amssymb,superscriptaddress, 10pt]{revtex4-2}

\usepackage[usenames,dvipsnames]{xcolor}
\usepackage{amsmath,amssymb}
\usepackage{graphicx}
\usepackage{hyperref}
\usepackage{braket}
\usepackage[normalem]{ulem}
\usepackage{enumerate}
\usepackage{MnSymbol}
\usepackage{esint}
\usepackage{comment}

\usepackage[normalem]{ulem}

\newcommand{\be}{\begin{equation}}
\newcommand{\ee}{\end{equation}}
\def\bes{\begin{subequations}}
\def\esu{\end{subequations}}

\newcommand{\dr}{\text{dr}}

\newcommand{\eff}{\text{eff}}

\newcommand{\dd}{{\rm d}}
\newcommand{\g}{g_\text{1D}}

\newcommand{\para}[1]{ \noindent\emph{\textbf{#1---}}}

\newcommand{\wind}{\mathcal{W}}

\begin{document}

\newcommand{\titleinfo}{Exotic critical states as fractional Fermi seas in the one-dimensional Bose gas}
\title{\titleinfo}
\author{Alvise Bastianello}\thanks{These authors contributed equally to this work.}
\affiliation{CEREMADE, CNRS, Universit\'e Paris-Dauphine, Universit\'e PSL, 75016 Paris, France}
\email{alvise.bastianello@dauphine.psl.eu}

\author{Yi Zeng}\thanks{These authors contributed equally to this work.}
\affiliation{Institut f{\"u}r Experimentalphysik und Zentrum f{\"u}r Quantenphysik, Universit{\"a}t Innsbruck, Technikerstra{\ss}e 25, Innsbruck, 6020, Austria}

\author{Sudipta Dhar}
\affiliation{Institut f{\"u}r Experimentalphysik und Zentrum f{\"u}r Quantenphysik, Universit{\"a}t Innsbruck, Technikerstra{\ss}e 25, Innsbruck, 6020, Austria}
\author{Zekui Wang}
\affiliation{Institut f{\"u}r Experimentalphysik und Zentrum f{\"u}r Quantenphysik, Universit{\"a}t Innsbruck, Technikerstra{\ss}e 25, Innsbruck, 6020, Austria}
\affiliation{
State Key Laboratory of Quantum Optics Technologies and Devices, Institute of Opto-Electronics, Shanxi University, Taiyuan 030006, P. R. China}

\author{Xudong Yu}
\affiliation{Institut f{\"u}r Experimentalphysik und Zentrum f{\"u}r Quantenphysik, Universit{\"a}t Innsbruck, Technikerstra{\ss}e 25, Innsbruck, 6020, Austria}

\author{Milena Horvath}
\affiliation{Institut f{\"u}r Experimentalphysik und Zentrum f{\"u}r Quantenphysik, Universit{\"a}t Innsbruck, Technikerstra{\ss}e 25, Innsbruck, 6020, Austria}

\author{Grigori E. Astrakharchik}
\affiliation{Departament de Física, Campus Nord B4-B5, Universitat Politècnica de Catalunya, E-08034 Barcelona, Spain}

\author{Yanliang Guo}
\email{Yanliang.Guo@uibk.ac.at}
\affiliation{Key Laboratory of Quantum State Construction and Manipulation (Ministry of Education), School of Physics, Renmin University of China, Beijing 100872, China}
\affiliation{Institut f{\"u}r Experimentalphysik und Zentrum f{\"u}r Quantenphysik, Universit{\"a}t Innsbruck, Technikerstra{\ss}e 25, Innsbruck, 6020, Austria}

\author{Hanns-Christoph  N{\"a}gerl}
\affiliation{Institut f{\"u}r Experimentalphysik und Zentrum f{\"u}r Quantenphysik, Universit{\"a}t Innsbruck, Technikerstra{\ss}e 25, Innsbruck, 6020, Austria}

\author{Manuele Landini}
\affiliation{Institut f{\"u}r Experimentalphysik und Zentrum f{\"u}r Quantenphysik, Universit{\"a}t Innsbruck, Technikerstra{\ss}e 25, Innsbruck, 6020, Austria}

\begin{abstract} Critical quantum field theories occupy a central position in modern theoretical physics for their inherent universality stemming from long-range correlations. As an example, the Tomonaga–Luttinger liquid (TLL) describes a wealth of one-dimensional quantum systems at low temperatures. Its behavior is deeply rooted in the emergence of an effective Fermi sea, leading to power-law correlations and Friedel oscillations. A promising direction to realize systems exhibiting novel universal behavior beyond TLL is through the generalization of the underlying Fermi sea. In this Letter, we show that fractional Fermi seas with reduced occupancy arise in an integrable Bose gas driven out of equilibrium by cyclic changes in interactions across the full range of repulsive and attractive regimes. 
The correlation functions feature signatures of criticality incompatible with a conventional TLL, suggesting a novel critical phase. Our predictions, based on Generalized Hydrodynamics, are directly relevant to cold atoms.
\end{abstract}
\maketitle

\para{Introduction.}  Low-temperature quantum matter manifests emergent universality deeply connected with the statistics and symmetries of its microscopic constituents, captured by effective quantum field theories~\cite{fradkin2013field}. Particularly important are critical, or gapless, field theories, featuring long-range correlations and possibly conformal invariance~\cite{francesco2012conformal}. A prominent example is the celebrated Tomonaga-Luttinger liquid (TLL)~\cite{Haldane1981,Cazalilla2004}, which describes the low-energy properties of a wealth of one dimensional quantum models. The TLL is associated with Fermi seas and low-energy excitations around the Fermi momentum $p_F$, resulting in correlation functions with power-law decay~\cite{Cazalilla2004} 
and Friedel oscillations (FO)~\cite{Cazalilla2004,Friedel1952}  with natural wavelength $\hbar/p_F$. While Fermi seas generally arise in fermionic systems, in one dimension quantum statistics and interactions are deeply intertwined. Hence, Fermi seas and TLL behavior may also emerge for interacting bosons at low-temperatures~\cite{vonDelft1998}.

A natural question is whether a generalization of Fermi seas is conceivable, still manifesting critical properties.
A possible pathway follows Haldane's proposal~\cite{Haldane1991,Wu2001} on generalized exclusion statistics (GES): Pauli exclusion principle is extended to allow at most $1/\alpha$ particles in the same state, interpolating between fermions ($\alpha=1$) and bosons ($\alpha=0$).
GES describes anyonic systems~\cite{Batchelor2006}, as well as bosons with tunable interactions~\cite{bernard1994,Batchelor2007},
bridging the non-interacting and hard-core (fermionized) regimes. 
Extending this idea, one can also ``hyperfermionize" the system with occupancy $1/\alpha<1$: we refer to these states where only one state every $\alpha$ is occupied as ``fractional Fermi seas" (FFS). FFS realized as the ground state of a system following GES are also described by TLL theory~\cite{Wu2001}.

\begin{figure}[b!]
\includegraphics[width=0.99\columnwidth]{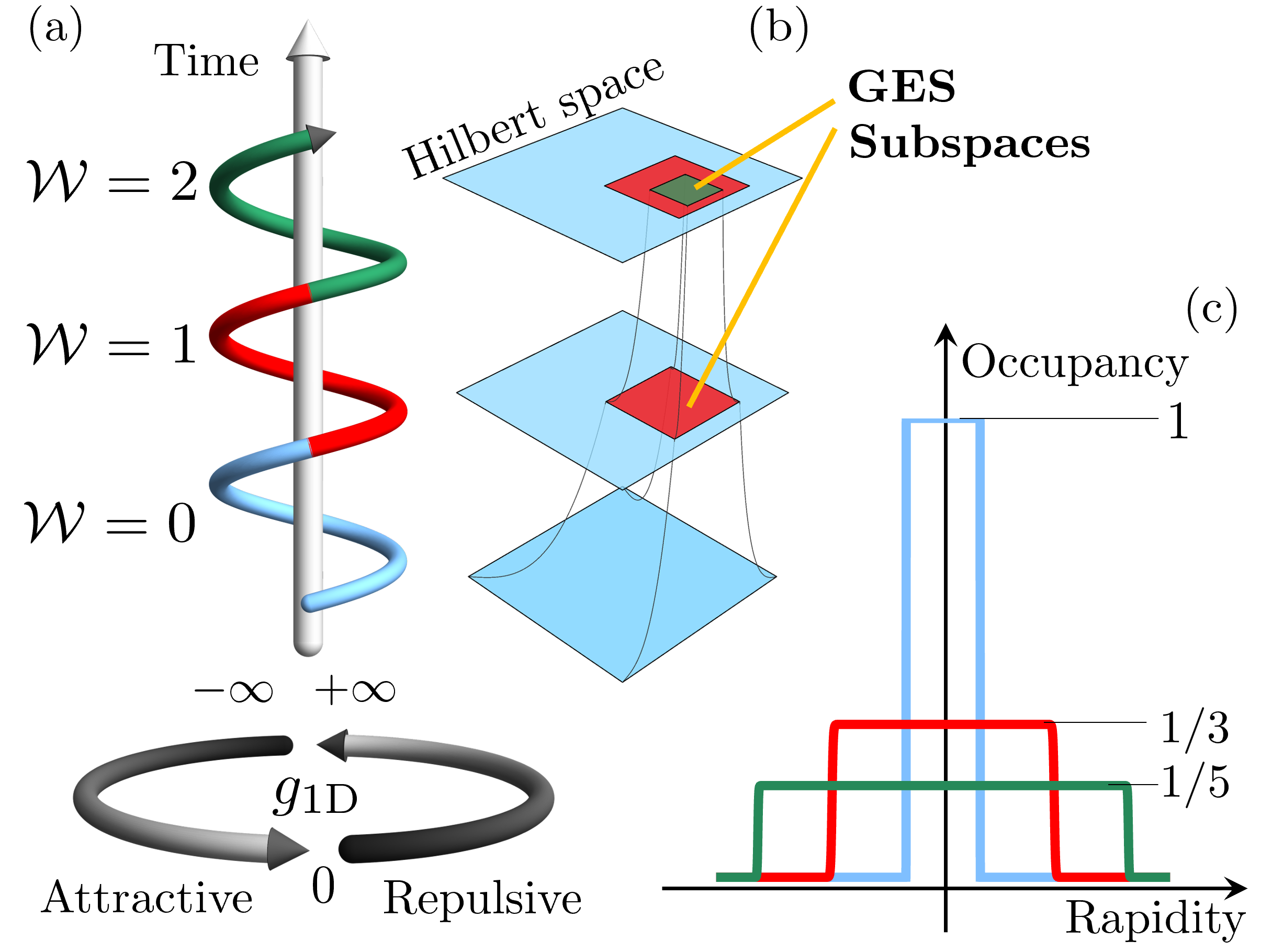}
\caption{\textbf{Interaction cycles and fractional Fermi seas.---}
(a) We perform slow cyclical changes of $\g$ crossing the TG-sTG transition. 
(b) Each cycle maps the initial GGE onto a GGE with reduced occupancy, realizing an analogue of a 
GES. 
(c) In particular, the initial ground state is transformed into a fractional Fermi sea.}
\label{Fig_1}
\end{figure}
In this Letter, we use interaction cycles~\cite{Yonezawa2013,Marciniak2025,Kao2021} in the integrable Lieb-Liniger (LL) model~\cite{Lieb1963} to realize FFS, and show that their correlation functions bear signatures of criticality beyond the conventional TLL. The experimental realization is presented in Ref.~\cite{ExpHolo}.
The LL model describes one-dimensional bosons with contact interactions with coupling $\g$~\cite{Olshanii1998}.
We consider nonequilibrium protocols where $\g$ is slowly and cyclically changed in time as depicted in Fig.~\ref{Fig_1}(a). Starting the protocol from weakly-repulsive values of $\g$, the coupling is ramped to strongly repulsive values, then quenched through the Tonks-Girardeau (TG) -- super Tonks-Girardeau (sTG) transition~\cite{Astrakharchik2005,Batchelor2005,Haller2009,Chen2010} to infinite attraction, and ultimately brought back to the non-interacting point $\g=0$. The cycle is repeated $\wind$ times.

In the cycle, FFS are not realized as ground states of a GES, but rather as highly-excited nonequilibrium states described by Generalized Gibbs Ensembles (GGE)~\cite{Rigol2007,Ilievski2015,Calabrese2016} built on the infinitely many conserved quantities of the LL model~\cite{Lieb1963}, which also implies the existence of stable quasiparticles parametrized by their rapidity $\lambda$~\cite{takahashi2005thermodynamics}. At finite density of excitations, quasiparticles become collective modes strongly renormalized by many-body interactions.
The rapidities generalize the notion of a single-particle wavevector of non-interacting systems to the interacting case. 
In such eigenstates in the thermodynamic limit, the occupancy $\vartheta(\lambda)$ describes the fraction of occupied states and it is valued between zero and one. In the ground state, $\vartheta(\lambda)$ is a maximally-occupied Fermi sea. 
We use Generalized Hydrodynamics (GHD)~\cite{Alvaredo2016,Bertini2016,Bastianello2022,Doyon2024}, a hydrodynamic approach to nearly-integrable models, to show that after the cycle is repeated $\wind$ times the initial GGE is mapped into a new one with reduced maximum occupancy $\vartheta(\lambda)\leq 1/(2\wind+1)$, in agreement with previous results obtained under the assumption of quantum adiabaticity (QA)~\cite{Yonezawa2013,Marciniak2025}. Within QA, the protocol is  also known as holonomy cycle~\cite{Yonezawa2013,Marciniak2025,Kao2021}: we choose a separate notation to emphasize the difference with the hydrodynamic framework, as we later discuss. 
The cycle acts as a \emph{projector} onto a Hilbert subspace 
with an occupancy described by an effective fractional GES with $\alpha=2\wind+1$, see Fig.~\ref{Fig_1}(b). 
The initial ground state is mapped into an FFS, see Fig.~\ref{Fig_1}(c).

We show that FFS manifest signatures of criticality in the one-particle correlation function $g^{(1)}(x)\equiv \langle \hat{\psi}^\dagger(x)\hat{\psi}(0)\rangle/n$, where $n$ is the particle density and $\hat{\psi}$ is the bosonic annihilation operator with commutator $[\hat{\psi}(x),\hat{\psi}^\dagger(y)]=\delta(x-y)$.
Specifically, $g^{(1)}(x)$ features power-law decay of correlation functions and FO for any finite repulsive interaction strength, in agreement with predictions in the limit of zero or infinite repulsion~\cite{Marciniak2025}. 
The FFS realized in this nonequilibrium setting offer a richer paradigm beyond the conventional TLL theory.

\para{Interaction cycles and GHD.} 
The Lieb-Liniger Hamiltonian is
\be\label{eq_LL_H}
\hat{H}=\int \dd x \left(\frac{\hbar^2}{2m}\partial_x\hat{\psi}^\dagger\partial_x\hat{\psi}+\frac{\g}{2}\hat{\psi}^\dagger\hat{\psi}^\dagger\hat{\psi}\hat{\psi}\right)\, ,
\ee
where $\g$ is the pairwise interaction coupling and $m$ is the atom's mass. 
GHD is valid for slow and weakly inhomogeneous protocols and it incorporates weak integrability-breaking terms~\cite{Bastianello2021}. It was employed in cold-atom experiments accounting for the effect of traps 
\cite{Schemmer2019,Malvania2021,Schuttelkopf2024,dubois2024}, time-dependent couplings~\cite{horvath2025} and other integrability-breaking terms~\cite{Cataldini2022,Moller2021,Yang2024Phantom}.

Here, we summarize the key concepts of GHD~\cite{takahashi2005thermodynamics}, leaving further details to the Supplementary Material (SM)~\cite{suppmat}. 
The simultaneous eigenstates of $\hat{H}$ and of all the other conserved quantities are labeled by a set of rapidities $|\{\lambda_i\}_{i=1}^N\rangle$, each describing a quasiparticle.
Quasiparticles experience a scattering phase $\Theta( \lambda)=-2\arctan(\tfrac{\hbar^2\lambda}{m\g})$, which affects the rapidity quantization through the non-linear Bethe equations~\cite{bethe1931theorie,Lieb1963}. 
For large system sizes $L\to+\infty$, in the thermodynamic limit one introduces macrostates $|\{\lambda_i\}_{i=1}^N\rangle\to \rho(\lambda)$~\cite{takahashi2005thermodynamics} where the root density $\rho(\lambda)$ is the coarse-grained rapidity's density of state, directly accessible in experiments~\cite{Wilson2020,Malvania2021,dubois2024,Li2023,Yang2024Phantom,horvath2025}.
Importantly, several microstates share the same root density, which identifies a GGE~\cite{Ilievski2016StrCh} with extensive entropy $\mathcal{S}[\rho]$~\cite{suppmat}.
From the thermodynamic limit of the Bethe equations, the total root density is introduced $\rho^t(\lambda)=\frac{1}{2\pi}-\int \tfrac{\dd\lambda'}{2\pi}\varphi(\lambda-\lambda')\rho(\lambda')$, with $\varphi(\lambda)=\partial_\lambda \Theta(\lambda)$.
For a given state, $\rho^t(\lambda)$ describes the total available phase space compatible with interactions.
As a result of this, the occupancy $\vartheta(\lambda)\equiv\rho(\lambda)/\rho^t(\lambda)$ represents the fraction of occupied states. In particular, the ground state of the repulsive ($\g>0$) Hamiltonian \eqref{eq_LL_H} has zero entropy, and its occupancy $\vartheta(\lambda)$ is a Fermi sea with occupation of one for $|\lambda|<\lambda_\text{F}$ and zero otherwise. The Fermi rapidity $\lambda_\text{F}$ generalizes the Fermi momentum to the rapidity space. For $\g=0$, the root density coincides with the bosonic momentum distribution, whereas in the TG regime, $\rho(\lambda)$ is the momentum distribution of the fermionic quasiparticles describing the hard core bosons~\cite{Girardeau1960,Kinoshita2004,Wilson2020}.

In general, computing $\vartheta(\lambda)$ in a non-equilibrium protocol is challenging~\cite{Caux2016,Piroli2017}, but for the case of slow changes in the coupling, it can be determined within GHD. 
Describing realistic experiments requires accounting for finite-temperature and inhomogeneities~\cite{ExpHolo}. Here, we consider homogeneous systems starting from the ground state of $\hat{H}$ for $\g>0$.
In the attractive regime $\g<0$, quasiparticles can form bound states~\cite{McGuire1964}.
At the TG-sTG transition, bound state formation is suppressed due to their divergent binding energy~\cite{Astrakharchik2005,Batchelor2005,Haller2009}, while at finite $\g$, the formation of bound states is halted by integrability~\cite{Piroli2015,Koch2021,horvath2025}. Therefore, when the cycle is traversed in the forward direction
no bound states are generated. 
In this case, the GHD equations reduce to~\cite{Bastianello2019}
\be\label{eq_ghd}
\partial_t\vartheta(\lambda)+a^\text{eff}(\lambda)\partial_\lambda \vartheta(\lambda)=0\, ,
\ee
where the effective acceleration $a^\eff(\lambda)\propto \partial_t \g$  is a non-linear function of the state itself, see SM~\cite{suppmat}.
\begin{figure*}[t!]
\includegraphics[width=0.99\textwidth]{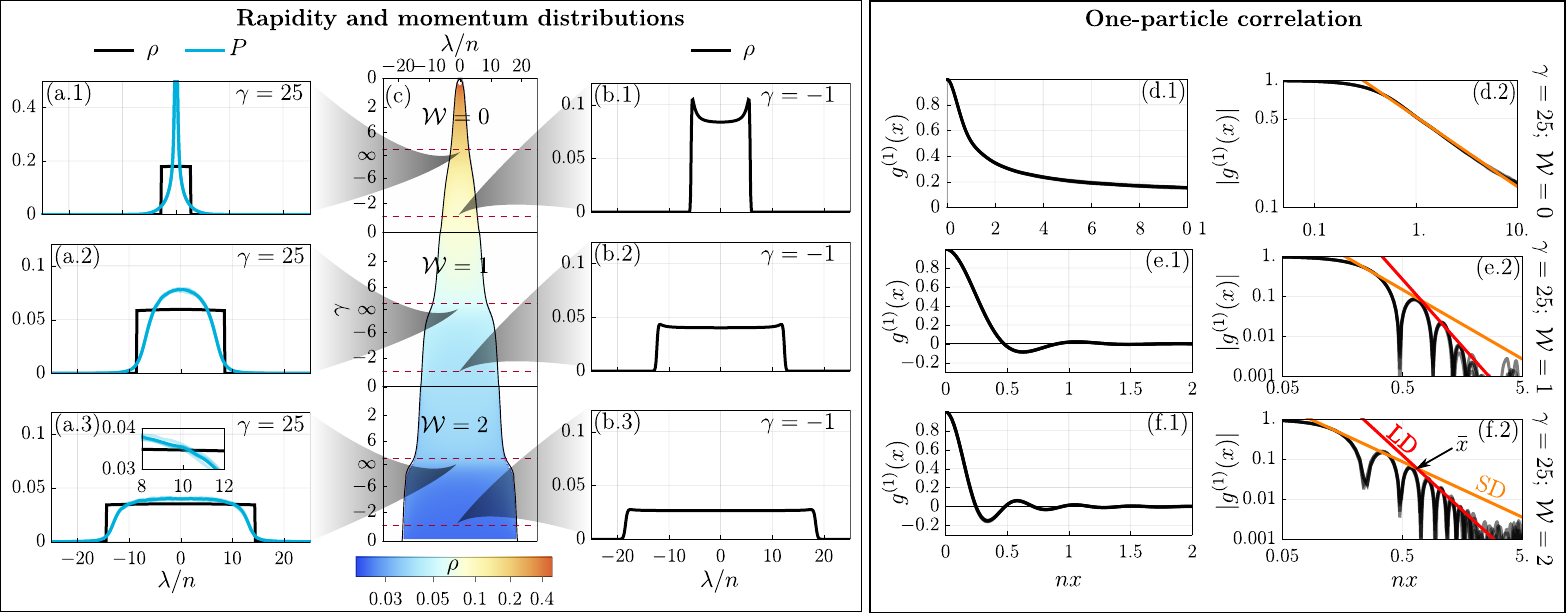}
\caption{\textbf{Rapidity and momentum distribution, and one-particle correlation.---}Panels (a,b,c). We plot the smooth evolution of the root density $\rho(\lambda)$ versus the normalized interaction strength $\gamma\equiv 2m\g/(\hbar^2 n)$ in the cycle, where the vertical axis is shown on the non-linear scale $(2/\pi)\arctan(\gamma/3)+(1-\text{sign}(\gamma))+2\wind$. Due to finite interaction strength, $\rho(\lambda)$ shows a $\gamma$- dependent concavity (more prominent at small $\gamma$).
In the repulsive regime (a.1,a.2,a.3), we also plot the momentum distribution $P(p)$ as a function of $\lambda\equiv p/\hbar$, to facilitate its comparison with $\rho(\lambda)$. The statistical error is negligible on the plot scale: in the inset of panel a.2, we show in light blue the partial averages obtained by dividing the total number of samples in five groups with the same size. The small relative distance shows the convergence of the Monte Carlo sampling,  see SM~\cite{suppmat} for details. Panels (d,e,f). Examples of $g^{(1)}(x)$ in linear (d.1,e.1,f.1) and log-log (d.2,e.2,f.2) scales, showing FO with a power-law decaying amplitude and a crossover between a short-distance (SD) and a faster long-distance (LD) power-law at a distance $\bar{x}$.
We show the case of $\gamma=25$, with a prominent power-law crossover. 
Gray lines are data of five independent Monte Carlo simulations and the black line is their average. Deviations from the average are appreciable only in the tails of the plots in log-log scale. The apparent large-distance plateau visible in the log-log scale is due to the statistical noise of the Monte Carlo simulations~\cite{suppmat}.
}\label{Fig_n2}
\end{figure*}
  The solution of Eq.~\eqref{eq_ghd} parametrically depends on time through $\g$ 
  and the branch of the cycle $\wind$, hence we parametrize the solution as $\vartheta_{\g,\wind}(\lambda)$.
Equation~\eqref{eq_ghd} is valid both in the repulsive and attractive regimes, and must be properly continued when joining the two at $\g=0$ and $\g=\pm\infty$, imposing the continuity of $\rho(\lambda)$~\cite{Koch2021}. At the TG-sTG transition $\rho^t(\lambda)$ approaches a constant value, hence the continuity of $\rho(\lambda)$ implies that $\vartheta(\lambda)$ is also continuous. 
On the other hand, at the zero-crossing (or $\g=0$ point) $\rho^t(\lambda)$ tends to $\tfrac{1}{2\pi}+\text{sign}(\g)\rho(\lambda)$,
leading to the requirement $1/\vartheta_{\g=0^+,\wind+1}=2+1/\vartheta_{\g=0^-,\wind}$.
With this condition, an implicit solution to Eq.~\eqref{eq_ghd} is given through the method of characteristics~\cite{Bastianello2019,Moller2020,suppmat}. Starting from the ground state, GHD predicts the announced FFS occupancies 
\be\label{eq_occ_FFS}
\vartheta_{\g,\wind}(\lambda)=\frac{1}{2\wind+1}\hspace{2pc}\text{if}\,\,\,\, |\lambda|<\lambda_\text{F}(\g,\wind)\, ,
\ee
and zero otherwise. 
Above, the running Fermi rapidity evolves according to Eq.~\eqref{eq_ghd} as $\partial_t\lambda_\text{F}(\g,\wind)=a^\eff(\lambda_\text{F}(\g,\wind))$. More conveniently, it can be fixed by the density, $n=\int \dd\lambda\,  \rho(\lambda)$. 
While Eq.~\eqref{eq_occ_FFS} describes the evolution from the ground state's GGE, the continuity conditions at $\g=0$ for the occupancy $\vartheta$ are always valid, and show that arbitrary initial GGEs are mapped into new GGEs with reduced maximum occupancy $\vartheta_{\g,\wind}(\lambda)\le (2\wind+1)^{-1}$ \cite{suppmat}, providing a non-equilibrium analogy with GES.
Previous mapping~\cite{bernard1994,Batchelor2007} based on $\rho(\lambda)$ showed that the equilibrium thermodynamics of integrable models equals the GES's thermodynamics~\cite{Wu2001} for a $\lambda-$dependent generalization of $0<\alpha<1$.

Here, instead, we suggest a far-from-equilibrium analogy with fractional GES with $\alpha>1$ on the occupancy $\vartheta(\lambda)$.
Yet, the correspondence is not exact, as the fractional statistics is here realized through conservation laws and not with a statistical constraint.  
As the conserved quantities, such as energy and number of particles, are linear functions of $\rho(\lambda)$, a suitably-chosen GES that in its ground state has the same root density of the FFS would have, by definition, the same conserved quantities.
However, operators that do not commute with the conserved charges can experience deviations visiting virtual states of the Hilbert space beyond the projected sector, see Fig.~\ref{Fig_1}(b). For example, this is the case for the $g^{(1)}(x)$ correlation function. As a GES's ground state is a maximally-occupied Fermi sea compatible with the statistics, it behaves as a nearly -rigid Fermi sea where only excitations around the Fermi edges contribute to low-energy modes, resulting in a conventional TLL description~\cite{Wu2001}.
In contrast, in the FFS realized as GGEs, the Pauli blockade is not enforced since the occupancy is not maximal. Hence, low-energy modes do not involve the Fermi edges only, but also the bulk.
Nonetheless, the sharp discontinuities in $\vartheta(\lambda)$ still leaves signature of criticality in $g^{(1)}(x)$ beyond conventional TLL~\cite{DeNardis2018}.
\begin{figure}[t!]
\includegraphics[width=0.99\columnwidth]{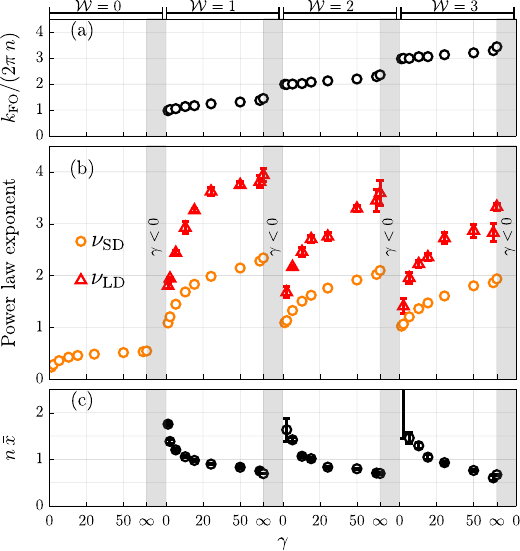}
\caption{\textbf{Power law exponent and FO in one-particle correlation function for $\gamma>0$--- }Panel (a) shows the FO's frequency, which is $2\pi n \wind$ at $\gamma=0$~\cite{Marciniak2025} and it deviates from it at finite $\gamma$, shown on the same nonlinear scale of Fig.~\ref{Fig_n2}. Panel (b). Short-distance (SD) and long-distance (LD) power law exponents, see Fig.~\ref{Fig_n2}(b). For details on the fit, see SM~\cite{suppmat}. Panel (c). Crossover distance $\bar{x}$ between the two power laws, defined as the intersection of the two power laws as in Fig.~\ref{Fig_n2}(b).
Error bars are obtained by comparing the fits of five independent Monte Carlo samplings for each value of $\gamma$ and $\wind$, and are omitted when negligible on the plot scale. We show the error bars only for the LD exponent in Panel (b) and $\bar{x}$ in Panel (a).
}   \label{Fig_n3}
\end{figure}

\para{Exotic criticality.} To investigate the emergent critical properties of the FFS, we focus on $g^{(1)}(x)$, as inspired by predictions in conventional TLL~\cite{Cazalilla2004} and motivated by its relevance for cold-atom experiments~\cite{ExpHolo}.
Exact analytical results are not available for $g^{(1)}(x)$ in the thermodynamic limit for finite $\g$ and at all distances~\cite{vonDelft1998,Cazalilla2004,Takacs2025,Wei2015}. However, for repulsive coupling $\g>0$, $g^{(1)}(x)$ can be computed via a Monte Carlo algorithm~\cite{senese2025} leveraging the Bethe ansatz results~\cite{Caux2007},  see also Refs.~\cite{suppmat,Alba2016,Dubail2020,Zhang2024}.
With this method, we compute the momentum distribution and then $g^{(1)}(x)$ through Fourier transform (Fig.~\ref{Fig_n2}). For $(\g>0,\wind=0)$, where conventional TLL holds, $g^{(1)}(x)$ shows a single power-law decay with no appreciable oscillations. This is expected, as TLL predicts $g_1(x)\simeq \text{(const)}\times|x|^{-\frac{1}{2K}}$ with $K$ being the Luttinger parameter, and FO are sub-leading in this regime. In the excited branches with $\wind>0$, $g^{(1)}(x)$ oscillates with a power-law decaying envelope at any finite interaction strength, consistent with Ref.~\cite{Marciniak2025} at the non-interacting and TG points.
Also, $g^{(1)}(x)$ shows other exotic properties, such as a crossover from a slower to a faster power law, see Fig.~\ref{Fig_n2}(d.2,e.2,f.2).
We fit the numerical data with the function $A \cos(k_\text{FO} x+\phi)/x^\nu$: while we found the FO wavevector $k_\text{FO}$ to be the same at all distances, we identify a transition between a short-distance powerlaw exponent $\nu_\text{SD}$ and a faster decay at long-distance governed by the exponent $\nu_\text{LD}$.
The crossover distance $\bar{x}$, defined as the distance where the short-distance and large-distance envelopes are equal~\cite{suppmat}, non-trivially changes with interactions, diverging at the non-interacting points where the analytic result $g^{(1)}(x)|_{g=0}=\sin( 2\pi n\wind x)/(2\pi n\wind x) $ holds~\cite{Marciniak2025}, and $\bar{x}$ moves to shorter distances as the interaction strength increases. As $\g$ increases, the decay becomes faster, eventually transitioning from a oscillating power law to an oscillating exponential decay in the TG regime~\cite{suppmat}.

Figure~\ref{Fig_n3} reports the observed power-law exponents and FO.
The oscillation frequency $k_\text{FO}$ (Fig.~\ref{Fig_n3}(a)) changes monotonically during the cycle. 
Within the TLL description, FO are expected at frequencies that are $2\pi$-multiples of the density.
For $\wind\ge 1$ we observe deviations, except at $\g=0$. 
In Fig.~\ref{Fig_n3}(b) we report the values of the power-law exponents, and the crossover distance between the slower and faster decay is shown in Fig.~\ref{Fig_n3}(c).
Within each branch, the exponents increase with interaction strength and the crossover distance decreases. For fixed $\g$, the exponent decreases for larger $\wind$. This is expected, since each cycle pumps energy into kinetic degrees of freedom, making interactions less relevant. 
The emergence of power-law correlations in a highly excited state is remarkable, and
we argue that the exotic behavior of $g^{(1)}(x)$, featuring dominant FO and bi-modal power law decay, is the signature of an underlying beyond-TLL critical field theory.  
\begin{figure}[b!]
\includegraphics[width=0.99\columnwidth]{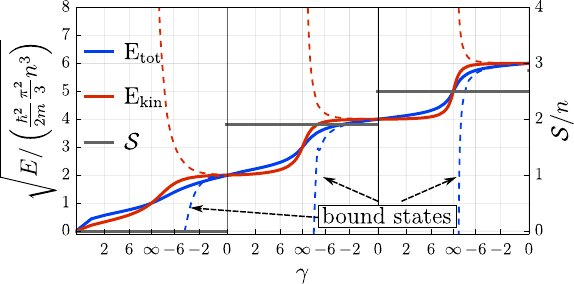}
\caption{\textbf{Energy and reversibility---} We show in blue the energy $E_\text{tot}$ and in red the kinetic energy $E_\text{kin}$ during the cycle. The normalized interaction strength (lower axis) is shown on a non-linear scale (same as Fig.~\ref{Fig_n2}).
Reversibility is broken at $\gamma=0$ as manifested in both the jump in entropy in the forward cycle and a strong variation in the energies ($E_\text{tot}$ dashed blue line, $E_\text{kin}$ dashed red line), due to the formation of bound states in the reversed cycle~\cite{Koch2021,horvath2025,suppmat}.
}\label{Fig_n4}
\end{figure}

\para{Comparing GHD with QA.} In
Fig.~\ref{Fig_n4} we show the total energy density $E=\langle \tfrac{\hbar^2}{2m}\partial_x\hat{\psi}^\dagger\partial_x\hat{\psi}+\tfrac{\g}{2}\hat{\psi}^\dagger\hat{\psi}^\dagger\hat{\psi}\hat{\psi}\rangle$ and kinetic energy density $E_\text{kin}=\langle \tfrac{\hbar^2}{2m}\partial_x\hat{\psi}^\dagger\partial_x\hat{\psi}\rangle$ predicted by GHD through Eq.~\eqref{eq_ghd}~\cite{suppmat}. These predictions coincide with those of QA~\cite{Marciniak2025}, naturally motivating its comparison with GHD. 
In the QA framework, the state evolves following a \emph{single eigenstate} of the instantaneous Hamiltonian, and its evolution is fully reversible. QA is expected to be valid for protocols acting on time scales larger than the inverse energy gap between the instantaneous eigenstate and the rest of the spectrum. This gap closes in the thermodynamic limit.
Within GHD, instead, the state evolves in a superposition of many instantaneous eigenstates (microstates) described by the same GGE, among which the eigenstate evolved within QA is found. 
Unlike QA, the GHD description remains valid also in the thermodynamic limit, with controlled expansions~\cite{Nardis2018,Bastianello2020,Durnin2021,Hubner2025} beyond the first-order derivatives of Eq. \eqref{eq_ghd}.

Eigenstates from the same GGE are indistinguishable based on local correlation functions~\cite{Caux2013,Essler2024}, hence the identical energy predictions. However, within GHD reversibility is broken. 
When considering the evolution in the reverse direction of the cycle, bound state formation cannot be ignored~\cite{horvath2025}. 
Whenever the $\g=0$ point is crossed, bound states are formed~\cite{Koch2021,Koch2022}, resulting in a strong energy increase, see Fig.~\ref{Fig_n4}.  
This ``staircase reversibility" is also captured by the macrostate's entropy $\mathcal{S}$ density~\cite{suppmat}
\be
\mathcal{S}=n[(1+2\wind)\log(1+2\wind)-(2\wind)\log(2\wind)]\, ,
\ee
which is constant within each branch $\wind$, and suddenly jumps crossing $\g=0$, signaling the irreversibility of the crossing, see Fig.~\ref{Fig_n4}.

\para{Discussion.} 
We have studied interaction cycles in the LL model as a way to engineer exotic far-from-equilibrium states described by FFS. These states are realized as GGEs, protected from thermalization by integrability, and predicted by GHD. 
We show how FFSs bear signatures of a novel critical phase in the one-particle correlation function, featuring prominent FO and power law decay at any repulsive interaction. 
A similar behavior is expected for FFS with attractive interactions.

A recent experiment~\cite{Kao2021} investigated interaction cycles in a dipolar Bose gas. Crucially, the dipolar interactions enhance the stability of the attractive regime~\cite{Chen2023}, usually unstable~\cite{Gerton2000,Donley2001,Strecker2002}, at the price of breaking integrability~\cite{Panfil2023,biagetti2025}.
In principle, non-integrable dynamics melts the FFS through relaxation of the GGE to a conventional finite-temperature Gibbs ensemble.
In a joint submission~\cite{ExpHolo}, we present the experimental realization of interactions cycles in ultracold cesium atoms more faithfully described by the LL model \eqref{eq_LL_H}, observing FO in $g^{(1)}(x)$ in quantitative agreement with the GHD framework developed in this Letter. 

The FFS we presented suggest an analogy with GES, which is here realized out of equilibrium by projecting initial states in subspaces of the Hilbert space with reduced occupancy. As the realization of GES only pertains to a sector of the Hilbert space, correlation functions involving virtual states beyond this sector feature exotic behavior beyond TLL predictions expected for conventional GES,
as we detect in the one-particle correlation.
Understanding how much the analogy with GES can be pursued, and developing a field-theory approach describing this novel critical phase, are among the most intriguing directions opened by this work.

\noindent\para{Data availability.} Data are available on Zenodo~\cite{Zenodo}.

\noindent\para{Note.}The additional references~\cite{Doyon2020,Gangardt2003,Doyon2017} appear in the supplementary material.

\noindent\para{Acknowledgments.}
The Innsbruck team acknowledges funding by the European Research Council (ERC) with project number 101201611 and by an FFG infrastructure grant with project number FO999896041. Y.Z. and  Y.G. are supported by the Austrian Academy of Sciences (\"OAW) with APART-MINT 12234 and by the Austrian Science Fund (FWF) with project number 10.55776/COE1, respectively. Z.W. is supported by the Quantum Science and Technology-National Science and Technology Major Project (Grant No. 2021ZD0302003) and the National Natural Science Foundation of China (Grant Nos. 12488301, 12034011, U23A6004). A.B. acknowledges partial support of the German Research Foundation (DFG) under the Germany's Excellence Strategy-EXC-2101-3990814868.
G. E. A. acknowledges financial support from Ministerio de Ciencia e Innovación MCIN/AEI/10.13039/501100011033 (Spain) under Grant No. PID2023-147469NB-C21.

\bibliography{biblio}

%apsrev4-2.bst 2019-01-14 (MD) hand-edited version of apsrev4-1.bst
%Control: key (0)
%Control: author (8) initials jnrlst
%Control: editor formatted (1) identically to author
%Control: production of article title (0) allowed
%Control: page (0) single
%Control: year (1) truncated
%Control: production of eprint (0) enabled
\begin{thebibliography}{77}%
\makeatletter
\providecommand \@ifxundefined [1]{%
 \@ifx{#1\undefined}
}%
\providecommand \@ifnum [1]{%
 \ifnum #1\expandafter \@firstoftwo
 \else \expandafter \@secondoftwo
 \fi
}%
\providecommand \@ifx [1]{%
 \ifx #1\expandafter \@firstoftwo
 \else \expandafter \@secondoftwo
 \fi
}%
\providecommand \natexlab [1]{#1}%
\providecommand \enquote  [1]{``#1''}%
\providecommand \bibnamefont  [1]{#1}%
\providecommand \bibfnamefont [1]{#1}%
\providecommand \citenamefont [1]{#1}%
\providecommand \href@noop [0]{\@secondoftwo}%
\providecommand \href [0]{\begingroup \@sanitize@url \@href}%
\providecommand \@href[1]{\@@startlink{#1}\@@href}%
\providecommand \@@href[1]{\endgroup#1\@@endlink}%
\providecommand \@sanitize@url [0]{\catcode `\\12\catcode `\$12\catcode
  `\&12\catcode `\#12\catcode `\^12\catcode `\_12\catcode `\%12\relax}%
\providecommand \@@startlink[1]{}%
\providecommand \@@endlink[0]{}%
\providecommand \url  [0]{\begingroup\@sanitize@url \@url }%
\providecommand \@url [1]{\endgroup\@href {#1}{\urlprefix }}%
\providecommand \urlprefix  [0]{URL }%
\providecommand \Eprint [0]{\href }%
\providecommand \doibase [0]{https://doi.org/}%
\providecommand \selectlanguage [0]{\@gobble}%
\providecommand \bibinfo  [0]{\@secondoftwo}%
\providecommand \bibfield  [0]{\@secondoftwo}%
\providecommand \translation [1]{[#1]}%
\providecommand \BibitemOpen [0]{}%
\providecommand \bibitemStop [0]{}%
\providecommand \bibitemNoStop [0]{.\EOS\space}%
\providecommand \EOS [0]{\spacefactor3000\relax}%
\providecommand \BibitemShut  [1]{\csname bibitem#1\endcsname}%
\let\auto@bib@innerbib\@empty
%</preamble>
\bibitem [{\citenamefont {Fradkin}(2013)}]{fradkin2013field}%
  \BibitemOpen
  \bibfield  {author} {\bibinfo {author} {\bibfnamefont {E.}~\bibnamefont
  {Fradkin}},\ }\href@noop {} {\emph {\bibinfo {title} {Field theories of
  condensed matter physics}}}\ (\bibinfo  {publisher} {Cambridge University
  Press},\ \bibinfo {year} {2013})\BibitemShut {NoStop}%
\bibitem [{\citenamefont {Francesco}\ \emph {et~al.}(2012)\citenamefont
  {Francesco}, \citenamefont {Mathieu},\ and\ \citenamefont
  {S{\'e}n{\'e}chal}}]{francesco2012conformal}%
  \BibitemOpen
  \bibfield  {author} {\bibinfo {author} {\bibfnamefont {P.}~\bibnamefont
  {Francesco}}, \bibinfo {author} {\bibfnamefont {P.}~\bibnamefont {Mathieu}},\
  and\ \bibinfo {author} {\bibfnamefont {D.}~\bibnamefont {S{\'e}n{\'e}chal}},\
  }\href@noop {} {\emph {\bibinfo {title} {Conformal field theory}}}\ (\bibinfo
   {publisher} {Springer Science \& Business Media},\ \bibinfo {year}
  {2012})\BibitemShut {NoStop}%
\bibitem [{\citenamefont {Haldane}(1981)}]{Haldane1981}%
  \BibitemOpen
  \bibfield  {author} {\bibinfo {author} {\bibfnamefont {F.~D.~M.}\
  \bibnamefont {Haldane}},\ }\bibfield  {title} {\bibinfo {title} {Luttinger
  liquid theory' of one-dimensional quantum fluids. i. properties of the
  {L}uttinger model and their extension to the general 1d interacting spinless
  {F}ermi gas},\ }\href {https://doi.org/10.1088/0022-3719/14/19/010}
  {\bibfield  {journal} {\bibinfo  {journal} {Journal of Physics C: Solid State
  Physics}\ }\textbf {\bibinfo {volume} {14}},\ \bibinfo {pages} {2585}
  (\bibinfo {year} {1981})}\BibitemShut {NoStop}%
\bibitem [{\citenamefont {Cazalilla}(2004)}]{Cazalilla2004}%
  \BibitemOpen
  \bibfield  {author} {\bibinfo {author} {\bibfnamefont {M.~A.}\ \bibnamefont
  {Cazalilla}},\ }\bibfield  {title} {\bibinfo {title} {Bosonizing
  one-dimensional cold atomic gases},\ }\href
  {https://doi.org/10.1088/0953-4075/37/7/051} {\bibfield  {journal} {\bibinfo
  {journal} {Journal of Physics B: Atomic, Molecular and Optical Physics}\
  }\textbf {\bibinfo {volume} {37}},\ \bibinfo {pages} {S1} (\bibinfo {year}
  {2004})}\BibitemShut {NoStop}%
\bibitem [{\citenamefont {Friedel}(1952)}]{Friedel1952}%
  \BibitemOpen
  \bibfield  {author} {\bibinfo {author} {\bibfnamefont {J.}~\bibnamefont
  {Friedel}},\ }\bibfield  {title} {\bibinfo {title} {The distribution of
  electrons round impurities in monovalent metals},\ }\href
  {https://doi.org/10.1080/14786440208561086} {\bibfield  {journal} {\bibinfo
  {journal} {The London, Edinburgh, and Dublin Philosophical Magazine and
  Journal of Science}\ }\textbf {\bibinfo {volume} {43}},\ \bibinfo {pages}
  {153} (\bibinfo {year} {1952})}\BibitemShut {NoStop}%
\bibitem [{\citenamefont {von Delft}\ and\ \citenamefont
  {Schoeller}(1998)}]{vonDelft1998}%
  \BibitemOpen
  \bibfield  {author} {\bibinfo {author} {\bibfnamefont {J.}~\bibnamefont {von
  Delft}}\ and\ \bibinfo {author} {\bibfnamefont {H.}~\bibnamefont
  {Schoeller}},\ }\bibfield  {title} {\bibinfo {title} {Bosonization for
  beginners --- refermionization for experts},\ }\href
  {https://doi.org/10.1002/andp.19985100401} {\bibfield  {journal} {\bibinfo
  {journal} {Annalen der Physik}\ }\textbf {\bibinfo {volume} {510}},\ \bibinfo
  {pages} {225} (\bibinfo {year} {1998})}\BibitemShut {NoStop}%
\bibitem [{\citenamefont {Haldane}(1991)}]{Haldane1991}%
  \BibitemOpen
  \bibfield  {author} {\bibinfo {author} {\bibfnamefont {F.~D.~M.}\
  \bibnamefont {Haldane}},\ }\bibfield  {title} {\bibinfo {title}
  {``{F}ractional statistics'' in arbitrary dimensions: {a} generalization of
  the {P}auli principle},\ }\href {https://doi.org/10.1103/PhysRevLett.67.937}
  {\bibfield  {journal} {\bibinfo  {journal} {Phys. Rev. Lett.}\ }\textbf
  {\bibinfo {volume} {67}},\ \bibinfo {pages} {937} (\bibinfo {year}
  {1991})}\BibitemShut {NoStop}%
\bibitem [{\citenamefont {Wu}\ \emph {et~al.}(2001)\citenamefont {Wu},
  \citenamefont {Yu},\ and\ \citenamefont {Yang}}]{Wu2001}%
  \BibitemOpen
  \bibfield  {author} {\bibinfo {author} {\bibfnamefont {Y.-S.}\ \bibnamefont
  {Wu}}, \bibinfo {author} {\bibfnamefont {Y.}~\bibnamefont {Yu}},\ and\
  \bibinfo {author} {\bibfnamefont {H.-X.}\ \bibnamefont {Yang}},\ }\bibfield
  {title} {\bibinfo {title} {Characterization of one-dimensional {L}uttinger
  liquids in terms of fractional exclusion statistics},\ }\href
  {https://www.sciencedirect.com/science/article/pii/S0550321301001201}
  {\bibfield  {journal} {\bibinfo  {journal} {Nuclear Physics B}\ }\textbf
  {\bibinfo {volume} {604}},\ \bibinfo {pages} {551} (\bibinfo {year}
  {2001})}\BibitemShut {NoStop}%
\bibitem [{\citenamefont {Batchelor}\ and\ \citenamefont
  {Guan}(2006{\natexlab{a}})}]{Batchelor2006}%
  \BibitemOpen
  \bibfield  {author} {\bibinfo {author} {\bibfnamefont {M.~T.}\ \bibnamefont
  {Batchelor}}\ and\ \bibinfo {author} {\bibfnamefont {X.-W.}\ \bibnamefont
  {Guan}},\ }\bibfield  {title} {\bibinfo {title} {Generalized exclusion
  statistics and degenerate signature of strongly interacting anyons},\ }\href
  {https://doi.org/10.1103/PhysRevB.74.195121} {\bibfield  {journal} {\bibinfo
  {journal} {Phys. Rev. B}\ }\textbf {\bibinfo {volume} {74}},\ \bibinfo
  {pages} {195121} (\bibinfo {year} {2006}{\natexlab{a}})}\BibitemShut
  {NoStop}%
\bibitem [{\citenamefont {Bernard}\ and\ \citenamefont
  {Wu}(1994)}]{bernard1994}%
  \BibitemOpen
  \bibfield  {author} {\bibinfo {author} {\bibfnamefont {D.}~\bibnamefont
  {Bernard}}\ and\ \bibinfo {author} {\bibfnamefont {Y.-S.}\ \bibnamefont
  {Wu}},\ }\href {https://arxiv.org/abs/cond-mat/9404025} {\bibinfo {title} {A
  note on statistical interactions and the thermodynamic {B}ethe ansatz}}
  (\bibinfo {year} {1994}),\ \Eprint {https://arxiv.org/abs/9404025}
  {arXiv:9404025} \BibitemShut {NoStop}%
\bibitem [{\citenamefont {Batchelor}\ and\ \citenamefont
  {Guan}(2006{\natexlab{b}})}]{Batchelor2007}%
  \BibitemOpen
  \bibfield  {author} {\bibinfo {author} {\bibfnamefont {M.~T.}\ \bibnamefont
  {Batchelor}}\ and\ \bibinfo {author} {\bibfnamefont {X.-W.}\ \bibnamefont
  {Guan}},\ }\bibfield  {title} {\bibinfo {title} {Fermionization and
  fractional statistics in the strongly interacting one-dimensional {B}ose
  gas},\ }\href {https://doi.org/10.1002/lapl.2006100681} {\bibfield  {journal}
  {\bibinfo  {journal} {Laser Physics Letters}\ }\textbf {\bibinfo {volume}
  {4}},\ \bibinfo {pages} {77} (\bibinfo {year}
  {2006}{\natexlab{b}})}\BibitemShut {NoStop}%
\bibitem [{\citenamefont {Yonezawa}\ \emph {et~al.}(2013)\citenamefont
  {Yonezawa}, \citenamefont {Tanaka},\ and\ \citenamefont
  {Cheon}}]{Yonezawa2013}%
  \BibitemOpen
  \bibfield  {author} {\bibinfo {author} {\bibfnamefont {N.}~\bibnamefont
  {Yonezawa}}, \bibinfo {author} {\bibfnamefont {A.}~\bibnamefont {Tanaka}},\
  and\ \bibinfo {author} {\bibfnamefont {T.}~\bibnamefont {Cheon}},\ }\bibfield
   {title} {\bibinfo {title} {Quantum holonomy in the {L}ieb-{L}iniger model},\
  }\href {https://doi.org/10.1103/PhysRevA.87.062113} {\bibfield  {journal}
  {\bibinfo  {journal} {Phys. Rev. A}\ }\textbf {\bibinfo {volume} {87}},\
  \bibinfo {pages} {062113} (\bibinfo {year} {2013})}\BibitemShut {NoStop}%
\bibitem [{\citenamefont {Marciniak}\ \emph {et~al.}(2025)\citenamefont
  {Marciniak}, \citenamefont {Astrakharchik}, \citenamefont {Pawłowski},\ and\
  \citenamefont {Juliá-Díaz}}]{Marciniak2025}%
  \BibitemOpen
  \bibfield  {author} {\bibinfo {author} {\bibfnamefont {M.}~\bibnamefont
  {Marciniak}}, \bibinfo {author} {\bibfnamefont {G.~E.}\ \bibnamefont
  {Astrakharchik}}, \bibinfo {author} {\bibfnamefont {K.}~\bibnamefont
  {Pawłowski}},\ and\ \bibinfo {author} {\bibfnamefont {B.}~\bibnamefont
  {Juliá-Díaz}},\ }\href {https://arxiv.org/abs/2504.19569} {\bibinfo {title}
  {Fermionizing the ideal {B}ose gas via topological pumping}} (\bibinfo {year}
  {2025}),\ \Eprint {https://arxiv.org/abs/2504.19569} {arXiv:2504.19569}
  \BibitemShut {NoStop}%
\bibitem [{\citenamefont {Kao}\ \emph {et~al.}(2021)\citenamefont {Kao},
  \citenamefont {Li}, \citenamefont {Lin}, \citenamefont {Gopalakrishnan},\
  and\ \citenamefont {Lev}}]{Kao2021}%
  \BibitemOpen
  \bibfield  {author} {\bibinfo {author} {\bibfnamefont {W.}~\bibnamefont
  {Kao}}, \bibinfo {author} {\bibfnamefont {K.-Y.}\ \bibnamefont {Li}},
  \bibinfo {author} {\bibfnamefont {K.-Y.}\ \bibnamefont {Lin}}, \bibinfo
  {author} {\bibfnamefont {S.}~\bibnamefont {Gopalakrishnan}},\ and\ \bibinfo
  {author} {\bibfnamefont {B.~L.}\ \bibnamefont {Lev}},\ }\bibfield  {title}
  {\bibinfo {title} {Topological pumping of a 1d dipolar gas into strongly
  correlated prethermal states},\ }\href
  {https://doi.org/10.1126/science.abb4928} {\bibfield  {journal} {\bibinfo
  {journal} {Science}\ }\textbf {\bibinfo {volume} {371}},\ \bibinfo {pages}
  {296} (\bibinfo {year} {2021})}\BibitemShut {NoStop}%
\bibitem [{\citenamefont {Lieb}\ and\ \citenamefont
  {Liniger}(1963)}]{Lieb1963}%
  \BibitemOpen
  \bibfield  {author} {\bibinfo {author} {\bibfnamefont {E.~H.}\ \bibnamefont
  {Lieb}}\ and\ \bibinfo {author} {\bibfnamefont {W.}~\bibnamefont {Liniger}},\
  }\bibfield  {title} {\bibinfo {title} {Exact analysis of an interacting
  {B}ose gas. {I}. the general solution and the ground state},\ }\href
  {https://doi.org/10.1103/PhysRev.130.1605} {\bibfield  {journal} {\bibinfo
  {journal} {Phys. Rev.}\ }\textbf {\bibinfo {volume} {130}},\ \bibinfo {pages}
  {1605} (\bibinfo {year} {1963})}\BibitemShut {NoStop}%
\bibitem [{\citenamefont {Zeng}\ \emph {et~al.}(2026)\citenamefont {Zeng},
  \citenamefont {Bastianello}, \citenamefont {Dhar}, \citenamefont {Wang},
  \citenamefont {Yu}, \citenamefont {Horvath}, \citenamefont {Astrakharchik},
  \citenamefont {Guo}, \citenamefont {Nägerl},\ and\ \citenamefont
  {Landini}}]{ExpHolo}%
  \BibitemOpen
  \bibfield  {author} {\bibinfo {author} {\bibfnamefont {Y.}~\bibnamefont
  {Zeng}}, \bibinfo {author} {\bibfnamefont {A.}~\bibnamefont {Bastianello}},
  \bibinfo {author} {\bibfnamefont {S.}~\bibnamefont {Dhar}}, \bibinfo {author}
  {\bibfnamefont {Z.}~\bibnamefont {Wang}}, \bibinfo {author} {\bibfnamefont
  {X.}~\bibnamefont {Yu}}, \bibinfo {author} {\bibfnamefont {M.}~\bibnamefont
  {Horvath}}, \bibinfo {author} {\bibfnamefont {G.~E.}\ \bibnamefont
  {Astrakharchik}}, \bibinfo {author} {\bibfnamefont {Y.}~\bibnamefont {Guo}},
  \bibinfo {author} {\bibfnamefont {H.-C.}\ \bibnamefont {Nägerl}},\ and\
  \bibinfo {author} {\bibfnamefont {M.}~\bibnamefont {Landini}},\ }\href
  {https://arxiv.org/abs/2602.17657} {\bibinfo {title} {Realization of
  fractional fermi seas}} (\bibinfo {year} {2026}),\ \Eprint
  {https://arxiv.org/abs/2602.17657} {arXiv:2602.17657 [cond-mat.quant-gas]}
  \BibitemShut {NoStop}%
\bibitem [{\citenamefont {Olshanii}(1998)}]{Olshanii1998}%
  \BibitemOpen
  \bibfield  {author} {\bibinfo {author} {\bibfnamefont {M.}~\bibnamefont
  {Olshanii}},\ }\bibfield  {title} {\bibinfo {title} {Atomic scattering in the
  presence of an external confinement and a gas of impenetrable bosons},\
  }\href {https://doi.org/10.1103/PhysRevLett.81.938} {\bibfield  {journal}
  {\bibinfo  {journal} {Phys. Rev. Lett.}\ }\textbf {\bibinfo {volume} {81}},\
  \bibinfo {pages} {938} (\bibinfo {year} {1998})}\BibitemShut {NoStop}%
\bibitem [{\citenamefont {Astrakharchik}\ \emph {et~al.}(2005)\citenamefont
  {Astrakharchik}, \citenamefont {Boronat}, \citenamefont {Casulleras},\ and\
  \citenamefont {Giorgini}}]{Astrakharchik2005}%
  \BibitemOpen
  \bibfield  {author} {\bibinfo {author} {\bibfnamefont {G.~E.}\ \bibnamefont
  {Astrakharchik}}, \bibinfo {author} {\bibfnamefont {J.}~\bibnamefont
  {Boronat}}, \bibinfo {author} {\bibfnamefont {J.}~\bibnamefont
  {Casulleras}},\ and\ \bibinfo {author} {\bibfnamefont {S.}~\bibnamefont
  {Giorgini}},\ }\bibfield  {title} {\bibinfo {title} {Beyond the
  {T}onks-{G}irardeau gas: Strongly correlated regime in quasi-one-dimensional
  {B}ose gases},\ }\href {https://doi.org/10.1103/PhysRevLett.95.190407}
  {\bibfield  {journal} {\bibinfo  {journal} {Phys. Rev. Lett.}\ }\textbf
  {\bibinfo {volume} {95}},\ \bibinfo {pages} {190407} (\bibinfo {year}
  {2005})}\BibitemShut {NoStop}%
\bibitem [{\citenamefont {Batchelor}\ \emph {et~al.}(2005)\citenamefont
  {Batchelor}, \citenamefont {Bortz}, \citenamefont {Guan},\ and\ \citenamefont
  {Oelkers}}]{Batchelor2005}%
  \BibitemOpen
  \bibfield  {author} {\bibinfo {author} {\bibfnamefont {M.~T.}\ \bibnamefont
  {Batchelor}}, \bibinfo {author} {\bibfnamefont {M.}~\bibnamefont {Bortz}},
  \bibinfo {author} {\bibfnamefont {X.~W.}\ \bibnamefont {Guan}},\ and\
  \bibinfo {author} {\bibfnamefont {N.}~\bibnamefont {Oelkers}},\ }\bibfield
  {title} {\bibinfo {title} {Evidence for the super {T}onks–{G}irardeau
  gas},\ }\href {https://doi.org/10.1088/1742-5468/2005/10/L10001} {\bibfield
  {journal} {\bibinfo  {journal} {Journal of Statistical Mechanics: Theory and
  Experiment}\ }\textbf {\bibinfo {volume} {2005}},\ \bibinfo {pages} {L10001}
  (\bibinfo {year} {2005})}\BibitemShut {NoStop}%
\bibitem [{\citenamefont {Haller}\ \emph {et~al.}(2009)\citenamefont {Haller},
  \citenamefont {Gustavsson}, \citenamefont {Mark}, \citenamefont {Danzl},
  \citenamefont {Hart}, \citenamefont {Pupillo},\ and\ \citenamefont
  {Nägerl}}]{Haller2009}%
  \BibitemOpen
  \bibfield  {author} {\bibinfo {author} {\bibfnamefont {E.}~\bibnamefont
  {Haller}}, \bibinfo {author} {\bibfnamefont {M.}~\bibnamefont {Gustavsson}},
  \bibinfo {author} {\bibfnamefont {M.~J.}\ \bibnamefont {Mark}}, \bibinfo
  {author} {\bibfnamefont {J.~G.}\ \bibnamefont {Danzl}}, \bibinfo {author}
  {\bibfnamefont {R.}~\bibnamefont {Hart}}, \bibinfo {author} {\bibfnamefont
  {G.}~\bibnamefont {Pupillo}},\ and\ \bibinfo {author} {\bibfnamefont {H.-C.}\
  \bibnamefont {Nägerl}},\ }\bibfield  {title} {\bibinfo {title} {Realization
  of an excited, strongly correlated quantum gas phase},\ }\href
  {https://doi.org/10.1126/science.1175850} {\bibfield  {journal} {\bibinfo
  {journal} {Science}\ }\textbf {\bibinfo {volume} {325}},\ \bibinfo {pages}
  {1224} (\bibinfo {year} {2009})}\BibitemShut {NoStop}%
\bibitem [{\citenamefont {Chen}\ \emph {et~al.}(2010)\citenamefont {Chen},
  \citenamefont {Guan}, \citenamefont {Yin}, \citenamefont {Hao},\ and\
  \citenamefont {Guan}}]{Chen2010}%
  \BibitemOpen
  \bibfield  {author} {\bibinfo {author} {\bibfnamefont {S.}~\bibnamefont
  {Chen}}, \bibinfo {author} {\bibfnamefont {L.}~\bibnamefont {Guan}}, \bibinfo
  {author} {\bibfnamefont {X.}~\bibnamefont {Yin}}, \bibinfo {author}
  {\bibfnamefont {Y.}~\bibnamefont {Hao}},\ and\ \bibinfo {author}
  {\bibfnamefont {X.-W.}\ \bibnamefont {Guan}},\ }\bibfield  {title} {\bibinfo
  {title} {Transition from a tonks-girardeau gas to a super-tonks-girardeau gas
  as an exact many-body dynamics problem},\ }\href
  {https://doi.org/10.1103/PhysRevA.81.031609} {\bibfield  {journal} {\bibinfo
  {journal} {Phys. Rev. A}\ }\textbf {\bibinfo {volume} {81}},\ \bibinfo
  {pages} {031609} (\bibinfo {year} {2010})}\BibitemShut {NoStop}%
\bibitem [{\citenamefont {Rigol}\ \emph {et~al.}(2007)\citenamefont {Rigol},
  \citenamefont {Dunjko}, \citenamefont {Yurovsky},\ and\ \citenamefont
  {Olshanii}}]{Rigol2007}%
  \BibitemOpen
  \bibfield  {author} {\bibinfo {author} {\bibfnamefont {M.}~\bibnamefont
  {Rigol}}, \bibinfo {author} {\bibfnamefont {V.}~\bibnamefont {Dunjko}},
  \bibinfo {author} {\bibfnamefont {V.}~\bibnamefont {Yurovsky}},\ and\
  \bibinfo {author} {\bibfnamefont {M.}~\bibnamefont {Olshanii}},\ }\bibfield
  {title} {\bibinfo {title} {Relaxation in a completely integrable many-body
  quantum system: an ab initio study of the dynamics of the highly excited
  states of 1d lattice hard-core bosons},\ }\href
  {https://doi.org/10.1103/PhysRevLett.98.050405} {\bibfield  {journal}
  {\bibinfo  {journal} {Phys. Rev. Lett.}\ }\textbf {\bibinfo {volume} {98}},\
  \bibinfo {pages} {050405} (\bibinfo {year} {2007})}\BibitemShut {NoStop}%
\bibitem [{\citenamefont {Ilievski}\ \emph {et~al.}(2015)\citenamefont
  {Ilievski}, \citenamefont {De~Nardis}, \citenamefont {Wouters}, \citenamefont
  {Caux}, \citenamefont {Essler},\ and\ \citenamefont {Prosen}}]{Ilievski2015}%
  \BibitemOpen
  \bibfield  {author} {\bibinfo {author} {\bibfnamefont {E.}~\bibnamefont
  {Ilievski}}, \bibinfo {author} {\bibfnamefont {J.}~\bibnamefont {De~Nardis}},
  \bibinfo {author} {\bibfnamefont {B.}~\bibnamefont {Wouters}}, \bibinfo
  {author} {\bibfnamefont {J.-S.}\ \bibnamefont {Caux}}, \bibinfo {author}
  {\bibfnamefont {F.~H.~L.}\ \bibnamefont {Essler}},\ and\ \bibinfo {author}
  {\bibfnamefont {T.}~\bibnamefont {Prosen}},\ }\bibfield  {title} {\bibinfo
  {title} {Complete generalized {G}ibbs ensembles in an interacting theory},\
  }\href {https://doi.org/10.1103/PhysRevLett.115.157201} {\bibfield  {journal}
  {\bibinfo  {journal} {Phys. Rev. Lett.}\ }\textbf {\bibinfo {volume} {115}},\
  \bibinfo {pages} {157201} (\bibinfo {year} {2015})}\BibitemShut {NoStop}%
\bibitem [{\citenamefont {Calabrese}\ \emph {et~al.}(2016)\citenamefont
  {Calabrese}, \citenamefont {Essler},\ and\ \citenamefont
  {Mussardo}}]{Calabrese2016}%
  \BibitemOpen
  \bibfield  {author} {\bibinfo {author} {\bibfnamefont {P.}~\bibnamefont
  {Calabrese}}, \bibinfo {author} {\bibfnamefont {F.~H.~L.}\ \bibnamefont
  {Essler}},\ and\ \bibinfo {author} {\bibfnamefont {G.}~\bibnamefont
  {Mussardo}},\ }\bibfield  {title} {\bibinfo {title} {Introduction to
  ‘quantum integrability in out of equilibrium systems’},\ }\href
  {https://doi.org/10.1088/1742-5468/2016/06/064001} {\bibfield  {journal}
  {\bibinfo  {journal} {Journal of Statistical Mechanics: Theory and
  Experiment}\ }\textbf {\bibinfo {volume} {2016}},\ \bibinfo {pages} {064001}
  (\bibinfo {year} {2016})}\BibitemShut {NoStop}%
\bibitem [{\citenamefont {Takahashi}(2005)}]{takahashi2005thermodynamics}%
  \BibitemOpen
  \bibfield  {author} {\bibinfo {author} {\bibfnamefont {M.}~\bibnamefont
  {Takahashi}},\ }\href@noop {} {\emph {\bibinfo {title} {Thermodynamics of
  one-dimensional solvable models}}}\ (\bibinfo  {publisher} {Cambridge
  University Press},\ \bibinfo {year} {2005})\BibitemShut {NoStop}%
\bibitem [{\citenamefont {Castro-Alvaredo}\ \emph {et~al.}(2016)\citenamefont
  {Castro-Alvaredo}, \citenamefont {Doyon},\ and\ \citenamefont
  {Yoshimura}}]{Alvaredo2016}%
  \BibitemOpen
  \bibfield  {author} {\bibinfo {author} {\bibfnamefont {O.~A.}\ \bibnamefont
  {Castro-Alvaredo}}, \bibinfo {author} {\bibfnamefont {B.}~\bibnamefont
  {Doyon}},\ and\ \bibinfo {author} {\bibfnamefont {T.}~\bibnamefont
  {Yoshimura}},\ }\bibfield  {title} {\bibinfo {title} {Emergent hydrodynamics
  in integrable quantum systems out of equilibrium},\ }\href
  {https://doi.org/10.1103/PhysRevX.6.041065} {\bibfield  {journal} {\bibinfo
  {journal} {Phys. Rev. X}\ }\textbf {\bibinfo {volume} {6}},\ \bibinfo {pages}
  {041065} (\bibinfo {year} {2016})}\BibitemShut {NoStop}%
\bibitem [{\citenamefont {Bertini}\ \emph {et~al.}(2016)\citenamefont
  {Bertini}, \citenamefont {Collura}, \citenamefont {De~Nardis},\ and\
  \citenamefont {Fagotti}}]{Bertini2016}%
  \BibitemOpen
  \bibfield  {author} {\bibinfo {author} {\bibfnamefont {B.}~\bibnamefont
  {Bertini}}, \bibinfo {author} {\bibfnamefont {M.}~\bibnamefont {Collura}},
  \bibinfo {author} {\bibfnamefont {J.}~\bibnamefont {De~Nardis}},\ and\
  \bibinfo {author} {\bibfnamefont {M.}~\bibnamefont {Fagotti}},\ }\bibfield
  {title} {\bibinfo {title} {Transport in out-of-equilibrium xxz chains: Exact
  profiles of charges and currents},\ }\href
  {https://doi.org/10.1103/PhysRevLett.117.207201} {\bibfield  {journal}
  {\bibinfo  {journal} {Phys. Rev. Lett.}\ }\textbf {\bibinfo {volume} {117}},\
  \bibinfo {pages} {207201} (\bibinfo {year} {2016})}\BibitemShut {NoStop}%
\bibitem [{\citenamefont {Bastianello}\ \emph {et~al.}(2022)\citenamefont
  {Bastianello}, \citenamefont {Bertini}, \citenamefont {Doyon},\ and\
  \citenamefont {Vasseur}}]{Bastianello2022}%
  \BibitemOpen
  \bibfield  {author} {\bibinfo {author} {\bibfnamefont {A.}~\bibnamefont
  {Bastianello}}, \bibinfo {author} {\bibfnamefont {B.}~\bibnamefont
  {Bertini}}, \bibinfo {author} {\bibfnamefont {B.}~\bibnamefont {Doyon}},\
  and\ \bibinfo {author} {\bibfnamefont {R.}~\bibnamefont {Vasseur}},\
  }\bibfield  {title} {\bibinfo {title} {Introduction to the special issue on
  emergent hydrodynamics in integrable many-body systems},\ }\href
  {https://doi.org/10.1088/1742-5468/ac3e6a} {\bibfield  {journal} {\bibinfo
  {journal} {Journal of Statistical Mechanics: Theory and Experiment}\ }\textbf
  {\bibinfo {volume} {2022}},\ \bibinfo {pages} {014001} (\bibinfo {year}
  {2022})}\BibitemShut {NoStop}%
\bibitem [{\citenamefont {Doyon}\ \emph {et~al.}(2025)\citenamefont {Doyon},
  \citenamefont {Gopalakrishnan}, \citenamefont {M\o{}ller}, \citenamefont
  {Schmiedmayer},\ and\ \citenamefont {Vasseur}}]{Doyon2024}%
  \BibitemOpen
  \bibfield  {author} {\bibinfo {author} {\bibfnamefont {B.}~\bibnamefont
  {Doyon}}, \bibinfo {author} {\bibfnamefont {S.}~\bibnamefont
  {Gopalakrishnan}}, \bibinfo {author} {\bibfnamefont {F.}~\bibnamefont
  {M\o{}ller}}, \bibinfo {author} {\bibfnamefont {J.}~\bibnamefont
  {Schmiedmayer}},\ and\ \bibinfo {author} {\bibfnamefont {R.}~\bibnamefont
  {Vasseur}},\ }\bibfield  {title} {\bibinfo {title} {Generalized
  hydrodynamics: a perspective},\ }\href
  {https://doi.org/10.1103/PhysRevX.15.010501} {\bibfield  {journal} {\bibinfo
  {journal} {Phys. Rev. X}\ }\textbf {\bibinfo {volume} {15}},\ \bibinfo
  {pages} {010501} (\bibinfo {year} {2025})}\BibitemShut {NoStop}%
\bibitem [{\citenamefont {Bastianello}\ \emph {et~al.}(2021)\citenamefont
  {Bastianello}, \citenamefont {De~Luca},\ and\ \citenamefont
  {Vasseur}}]{Bastianello2021}%
  \BibitemOpen
  \bibfield  {author} {\bibinfo {author} {\bibfnamefont {A.}~\bibnamefont
  {Bastianello}}, \bibinfo {author} {\bibfnamefont {A.}~\bibnamefont
  {De~Luca}},\ and\ \bibinfo {author} {\bibfnamefont {R.}~\bibnamefont
  {Vasseur}},\ }\bibfield  {title} {\bibinfo {title} {Hydrodynamics of weak
  integrability breaking},\ }\href {https://doi.org/10.1088/1742-5468/ac26b2}
  {\bibfield  {journal} {\bibinfo  {journal} {Journal of Statistical Mechanics:
  Theory and Experiment}\ }\textbf {\bibinfo {volume} {2021}},\ \bibinfo
  {pages} {114003} (\bibinfo {year} {2021})}\BibitemShut {NoStop}%
\bibitem [{\citenamefont {Schemmer}\ \emph {et~al.}(2019)\citenamefont
  {Schemmer}, \citenamefont {Bouchoule}, \citenamefont {Doyon},\ and\
  \citenamefont {Dubail}}]{Schemmer2019}%
  \BibitemOpen
  \bibfield  {author} {\bibinfo {author} {\bibfnamefont {M.}~\bibnamefont
  {Schemmer}}, \bibinfo {author} {\bibfnamefont {I.}~\bibnamefont {Bouchoule}},
  \bibinfo {author} {\bibfnamefont {B.}~\bibnamefont {Doyon}},\ and\ \bibinfo
  {author} {\bibfnamefont {J.}~\bibnamefont {Dubail}},\ }\bibfield  {title}
  {\bibinfo {title} {Generalized hydrodynamics on an atom chip},\ }\href
  {https://doi.org/10.1103/PhysRevLett.122.090601} {\bibfield  {journal}
  {\bibinfo  {journal} {Phys. Rev. Lett.}\ }\textbf {\bibinfo {volume} {122}},\
  \bibinfo {pages} {090601} (\bibinfo {year} {2019})}\BibitemShut {NoStop}%
\bibitem [{\citenamefont {Malvania}\ \emph {et~al.}(2021)\citenamefont
  {Malvania}, \citenamefont {Zhang}, \citenamefont {Le}, \citenamefont
  {Dubail}, \citenamefont {Rigol},\ and\ \citenamefont {Weiss}}]{Malvania2021}%
  \BibitemOpen
  \bibfield  {author} {\bibinfo {author} {\bibfnamefont {N.}~\bibnamefont
  {Malvania}}, \bibinfo {author} {\bibfnamefont {Y.}~\bibnamefont {Zhang}},
  \bibinfo {author} {\bibfnamefont {Y.}~\bibnamefont {Le}}, \bibinfo {author}
  {\bibfnamefont {J.}~\bibnamefont {Dubail}}, \bibinfo {author} {\bibfnamefont
  {M.}~\bibnamefont {Rigol}},\ and\ \bibinfo {author} {\bibfnamefont {D.~S.}\
  \bibnamefont {Weiss}},\ }\bibfield  {title} {\bibinfo {title} {{Generalized
  hydrodynamics in strongly interacting 1D {B}ose gases}},\ }\href
  {https://doi.org/10.1126/science.abf0147} {\bibfield  {journal} {\bibinfo
  {journal} {Science}\ }\textbf {\bibinfo {volume} {373}},\ \bibinfo {pages}
  {1129} (\bibinfo {year} {2021})}\BibitemShut {NoStop}%
\bibitem [{\citenamefont {Schüttelkopf}\ \emph {et~al.}(2026)\citenamefont
  {Schüttelkopf}, \citenamefont {Tajik}, \citenamefont {Bazhan}, \citenamefont
  {Cataldini}, \citenamefont {Ji}, \citenamefont {Schmiedmayer},\ and\
  \citenamefont {Møller}}]{Schuttelkopf2024}%
  \BibitemOpen
  \bibfield  {author} {\bibinfo {author} {\bibfnamefont {P.}~\bibnamefont
  {Schüttelkopf}}, \bibinfo {author} {\bibfnamefont {M.}~\bibnamefont
  {Tajik}}, \bibinfo {author} {\bibfnamefont {N.}~\bibnamefont {Bazhan}},
  \bibinfo {author} {\bibfnamefont {F.}~\bibnamefont {Cataldini}}, \bibinfo
  {author} {\bibfnamefont {S.-C.}\ \bibnamefont {Ji}}, \bibinfo {author}
  {\bibfnamefont {J.}~\bibnamefont {Schmiedmayer}},\ and\ \bibinfo {author}
  {\bibfnamefont {F.}~\bibnamefont {Møller}},\ }\bibfield  {title} {\bibinfo
  {title} {Characterizing transport in a quantum gas by measuring {D}rude
  weights},\ }\href {https://doi.org/10.1126/science.ads8327} {\bibfield
  {journal} {\bibinfo  {journal} {Science}\ }\textbf {\bibinfo {volume}
  {391}},\ \bibinfo {pages} {290} (\bibinfo {year} {2026})}\BibitemShut
  {NoStop}%
\bibitem [{\citenamefont {Dubois}\ \emph {et~al.}(2024)\citenamefont {Dubois},
  \citenamefont {Th\'em\`eze}, \citenamefont {Nogrette}, \citenamefont
  {Dubail},\ and\ \citenamefont {Bouchoule}}]{dubois2024}%
  \BibitemOpen
  \bibfield  {author} {\bibinfo {author} {\bibfnamefont {L.}~\bibnamefont
  {Dubois}}, \bibinfo {author} {\bibfnamefont {G.}~\bibnamefont {Th\'em\`eze}},
  \bibinfo {author} {\bibfnamefont {F.}~\bibnamefont {Nogrette}}, \bibinfo
  {author} {\bibfnamefont {J.}~\bibnamefont {Dubail}},\ and\ \bibinfo {author}
  {\bibfnamefont {I.}~\bibnamefont {Bouchoule}},\ }\bibfield  {title} {\bibinfo
  {title} {Probing the local rapidity distribution of a one-dimensional {B}ose
  gas},\ }\href {https://doi.org/10.1103/PhysRevLett.133.113402} {\bibfield
  {journal} {\bibinfo  {journal} {Phys. Rev. Lett.}\ }\textbf {\bibinfo
  {volume} {133}},\ \bibinfo {pages} {113402} (\bibinfo {year}
  {2024})}\BibitemShut {NoStop}%
\bibitem [{\citenamefont {Horvath}\ \emph {et~al.}(2025)\citenamefont
  {Horvath}, \citenamefont {Bastianello}, \citenamefont {Dhar}, \citenamefont
  {Koch}, \citenamefont {Guo}, \citenamefont {Caux}, \citenamefont {Landini},\
  and\ \citenamefont {Nägerl}}]{horvath2025}%
  \BibitemOpen
  \bibfield  {author} {\bibinfo {author} {\bibfnamefont {M.}~\bibnamefont
  {Horvath}}, \bibinfo {author} {\bibfnamefont {A.}~\bibnamefont
  {Bastianello}}, \bibinfo {author} {\bibfnamefont {S.}~\bibnamefont {Dhar}},
  \bibinfo {author} {\bibfnamefont {R.}~\bibnamefont {Koch}}, \bibinfo {author}
  {\bibfnamefont {Y.}~\bibnamefont {Guo}}, \bibinfo {author} {\bibfnamefont
  {J.-S.}\ \bibnamefont {Caux}}, \bibinfo {author} {\bibfnamefont
  {M.}~\bibnamefont {Landini}},\ and\ \bibinfo {author} {\bibfnamefont {H.-C.}\
  \bibnamefont {Nägerl}},\ }\href {https://arxiv.org/abs/2505.10550} {\bibinfo
  {title} {Observing bethe strings in an attractive {B}ose gas far from
  equilibrium}} (\bibinfo {year} {2025}),\ \Eprint
  {https://arxiv.org/abs/2505.10550} {arXiv:2505.10550} \BibitemShut {NoStop}%
\bibitem [{\citenamefont {Cataldini}\ \emph {et~al.}(2022)\citenamefont
  {Cataldini}, \citenamefont {M\o{}ller}, \citenamefont {Tajik}, \citenamefont
  {Sabino}, \citenamefont {Ji}, \citenamefont {Mazets}, \citenamefont
  {Schweigler}, \citenamefont {Rauer},\ and\ \citenamefont
  {Schmiedmayer}}]{Cataldini2022}%
  \BibitemOpen
  \bibfield  {author} {\bibinfo {author} {\bibfnamefont {F.}~\bibnamefont
  {Cataldini}}, \bibinfo {author} {\bibfnamefont {F.}~\bibnamefont
  {M\o{}ller}}, \bibinfo {author} {\bibfnamefont {M.}~\bibnamefont {Tajik}},
  \bibinfo {author} {\bibfnamefont {J.~a.}\ \bibnamefont {Sabino}}, \bibinfo
  {author} {\bibfnamefont {S.-C.}\ \bibnamefont {Ji}}, \bibinfo {author}
  {\bibfnamefont {I.}~\bibnamefont {Mazets}}, \bibinfo {author} {\bibfnamefont
  {T.}~\bibnamefont {Schweigler}}, \bibinfo {author} {\bibfnamefont
  {B.}~\bibnamefont {Rauer}},\ and\ \bibinfo {author} {\bibfnamefont
  {J.}~\bibnamefont {Schmiedmayer}},\ }\bibfield  {title} {\bibinfo {title}
  {Emergent {P}auli blocking in a weakly interacting {B}ose gas},\ }\href
  {https://doi.org/10.1103/PhysRevX.12.041032} {\bibfield  {journal} {\bibinfo
  {journal} {Phys. Rev. X}\ }\textbf {\bibinfo {volume} {12}},\ \bibinfo
  {pages} {041032} (\bibinfo {year} {2022})}\BibitemShut {NoStop}%
\bibitem [{\citenamefont {M\o{}ller}\ \emph {et~al.}(2021)\citenamefont
  {M\o{}ller}, \citenamefont {Li}, \citenamefont {Mazets}, \citenamefont
  {Stimming}, \citenamefont {Zhou}, \citenamefont {Zhu}, \citenamefont {Chen},\
  and\ \citenamefont {Schmiedmayer}}]{Moller2021}%
  \BibitemOpen
  \bibfield  {author} {\bibinfo {author} {\bibfnamefont {F.}~\bibnamefont
  {M\o{}ller}}, \bibinfo {author} {\bibfnamefont {C.}~\bibnamefont {Li}},
  \bibinfo {author} {\bibfnamefont {I.}~\bibnamefont {Mazets}}, \bibinfo
  {author} {\bibfnamefont {H.-P.}\ \bibnamefont {Stimming}}, \bibinfo {author}
  {\bibfnamefont {T.}~\bibnamefont {Zhou}}, \bibinfo {author} {\bibfnamefont
  {Z.}~\bibnamefont {Zhu}}, \bibinfo {author} {\bibfnamefont {X.}~\bibnamefont
  {Chen}},\ and\ \bibinfo {author} {\bibfnamefont {J.}~\bibnamefont
  {Schmiedmayer}},\ }\bibfield  {title} {\bibinfo {title} {Extension of the
  generalized hydrodynamics to the dimensional crossover regime},\ }\href
  {https://doi.org/10.1103/PhysRevLett.126.090602} {\bibfield  {journal}
  {\bibinfo  {journal} {Phys. Rev. Lett.}\ }\textbf {\bibinfo {volume} {126}},\
  \bibinfo {pages} {090602} (\bibinfo {year} {2021})}\BibitemShut {NoStop}%
\bibitem [{\citenamefont {Yang}\ \emph {et~al.}(2024)\citenamefont {Yang},
  \citenamefont {Zhang}, \citenamefont {Li}, \citenamefont {Lin}, \citenamefont
  {Gopalakrishnan}, \citenamefont {Rigol},\ and\ \citenamefont
  {Lev}}]{Yang2024Phantom}%
  \BibitemOpen
  \bibfield  {author} {\bibinfo {author} {\bibfnamefont {K.}~\bibnamefont
  {Yang}}, \bibinfo {author} {\bibfnamefont {Y.}~\bibnamefont {Zhang}},
  \bibinfo {author} {\bibfnamefont {K.-Y.}\ \bibnamefont {Li}}, \bibinfo
  {author} {\bibfnamefont {K.-Y.}\ \bibnamefont {Lin}}, \bibinfo {author}
  {\bibfnamefont {S.}~\bibnamefont {Gopalakrishnan}}, \bibinfo {author}
  {\bibfnamefont {M.}~\bibnamefont {Rigol}},\ and\ \bibinfo {author}
  {\bibfnamefont {B.~L.}\ \bibnamefont {Lev}},\ }\bibfield  {title} {\bibinfo
  {title} {Phantom energy in the nonlinear response of a quantum many-body scar
  state},\ }\href {https://doi.org/10.1126/science.adk8978} {\bibfield
  {journal} {\bibinfo  {journal} {Science}\ }\textbf {\bibinfo {volume}
  {385}},\ \bibinfo {pages} {1063} (\bibinfo {year} {2024})}\BibitemShut
  {NoStop}%
\bibitem [{sup()}]{suppmat}%
  \BibitemOpen
  \href@noop {} {\bibinfo  {journal} {Supplementary Material}\ }\BibitemShut
  {NoStop}%
\bibitem [{\citenamefont {Bethe}(1931)}]{bethe1931theorie}%
  \BibitemOpen
\bibfield  {journal} {  }\bibfield  {author} {\bibinfo {author} {\bibfnamefont
  {H.}~\bibnamefont {Bethe}},\ }\bibfield  {title} {\bibinfo {title} {{Zur
  Theorie der Metalle}},\ }\href {https://doi.org/10.1007/BF01341708}
  {\bibfield  {journal} {\bibinfo  {journal} {Zeitschrift f{\"u}r Physik}\
  }\textbf {\bibinfo {volume} {71}},\ \bibinfo {pages} {205} (\bibinfo {year}
  {1931})}\BibitemShut {NoStop}%
\bibitem [{\citenamefont {Wilson}\ \emph {et~al.}(2020)\citenamefont {Wilson},
  \citenamefont {Malvania}, \citenamefont {Le}, \citenamefont {Zhang},
  \citenamefont {Rigol},\ and\ \citenamefont {Weiss}}]{Wilson2020}%
  \BibitemOpen
  \bibfield  {author} {\bibinfo {author} {\bibfnamefont {J.~M.}\ \bibnamefont
  {Wilson}}, \bibinfo {author} {\bibfnamefont {N.}~\bibnamefont {Malvania}},
  \bibinfo {author} {\bibfnamefont {Y.}~\bibnamefont {Le}}, \bibinfo {author}
  {\bibfnamefont {Y.}~\bibnamefont {Zhang}}, \bibinfo {author} {\bibfnamefont
  {M.}~\bibnamefont {Rigol}},\ and\ \bibinfo {author} {\bibfnamefont {D.~S.}\
  \bibnamefont {Weiss}},\ }\bibfield  {title} {\bibinfo {title} {Observation of
  dynamical fermionization},\ }\href {https://doi.org/10.1126/science.aaz0242}
  {\bibfield  {journal} {\bibinfo  {journal} {Science}\ }\textbf {\bibinfo
  {volume} {367}},\ \bibinfo {pages} {1461} (\bibinfo {year}
  {2020})}\BibitemShut {NoStop}%
\bibitem [{\citenamefont {Li}\ \emph {et~al.}(2023)\citenamefont {Li},
  \citenamefont {Zhang}, \citenamefont {Yang}, \citenamefont {Lin},
  \citenamefont {Gopalakrishnan}, \citenamefont {Rigol},\ and\ \citenamefont
  {Lev}}]{Li2023}%
  \BibitemOpen
  \bibfield  {author} {\bibinfo {author} {\bibfnamefont {K.-Y.}\ \bibnamefont
  {Li}}, \bibinfo {author} {\bibfnamefont {Y.}~\bibnamefont {Zhang}}, \bibinfo
  {author} {\bibfnamefont {K.}~\bibnamefont {Yang}}, \bibinfo {author}
  {\bibfnamefont {K.-Y.}\ \bibnamefont {Lin}}, \bibinfo {author} {\bibfnamefont
  {S.}~\bibnamefont {Gopalakrishnan}}, \bibinfo {author} {\bibfnamefont
  {M.}~\bibnamefont {Rigol}},\ and\ \bibinfo {author} {\bibfnamefont {B.~L.}\
  \bibnamefont {Lev}},\ }\bibfield  {title} {\bibinfo {title} {Rapidity and
  momentum distributions of one-dimensional dipolar quantum gases},\ }\href
  {https://doi.org/10.1103/PhysRevA.107.L061302} {\bibfield  {journal}
  {\bibinfo  {journal} {Phys. Rev. A}\ }\textbf {\bibinfo {volume} {107}},\
  \bibinfo {pages} {L061302} (\bibinfo {year} {2023})}\BibitemShut {NoStop}%
\bibitem [{\citenamefont {Ilievski}\ \emph {et~al.}(2016)\citenamefont
  {Ilievski}, \citenamefont {Quinn}, \citenamefont {De~Nardis},\ and\
  \citenamefont {Brockmann}}]{Ilievski2016StrCh}%
  \BibitemOpen
  \bibfield  {author} {\bibinfo {author} {\bibfnamefont {E.}~\bibnamefont
  {Ilievski}}, \bibinfo {author} {\bibfnamefont {E.}~\bibnamefont {Quinn}},
  \bibinfo {author} {\bibfnamefont {J.}~\bibnamefont {De~Nardis}},\ and\
  \bibinfo {author} {\bibfnamefont {M.}~\bibnamefont {Brockmann}},\ }\bibfield
  {title} {\bibinfo {title} {String-charge duality in integrable lattice
  models},\ }\href {https://doi.org/10.1088/1742-5468/2016/06/063101}
  {\bibfield  {journal} {\bibinfo  {journal} {Journal of Statistical Mechanics:
  Theory and Experiment}\ }\textbf {\bibinfo {volume} {2016}},\ \bibinfo
  {pages} {063101} (\bibinfo {year} {2016})}\BibitemShut {NoStop}%
\bibitem [{\citenamefont {Girardeau}(1960)}]{Girardeau1960}%
  \BibitemOpen
  \bibfield  {author} {\bibinfo {author} {\bibfnamefont {M.}~\bibnamefont
  {Girardeau}},\ }\bibfield  {title} {\bibinfo {title} {Relationship between
  systems of impenetrable bosons and fermions in one dimension},\ }\href
  {https://doi.org/10.1063/1.1703687} {\bibfield  {journal} {\bibinfo
  {journal} {Journal of Mathematical Physics}\ }\textbf {\bibinfo {volume}
  {1}},\ \bibinfo {pages} {516} (\bibinfo {year} {1960})}\BibitemShut {NoStop}%
\bibitem [{\citenamefont {Kinoshita}\ \emph {et~al.}(2004)\citenamefont
  {Kinoshita}, \citenamefont {Wenger},\ and\ \citenamefont
  {Weiss}}]{Kinoshita2004}%
  \BibitemOpen
  \bibfield  {author} {\bibinfo {author} {\bibfnamefont {T.}~\bibnamefont
  {Kinoshita}}, \bibinfo {author} {\bibfnamefont {T.}~\bibnamefont {Wenger}},\
  and\ \bibinfo {author} {\bibfnamefont {D.~S.}\ \bibnamefont {Weiss}},\
  }\bibfield  {title} {\bibinfo {title} {Observation of a one-dimensional
  {T}onks-{G}irardeau gas},\ }\href {https://doi.org/10.1126/science.1100700}
  {\bibfield  {journal} {\bibinfo  {journal} {Science}\ }\textbf {\bibinfo
  {volume} {305}},\ \bibinfo {pages} {1125} (\bibinfo {year}
  {2004})}\BibitemShut {NoStop}%
\bibitem [{\citenamefont {Caux}(2016)}]{Caux2016}%
  \BibitemOpen
  \bibfield  {author} {\bibinfo {author} {\bibfnamefont {J.-S.}\ \bibnamefont
  {Caux}},\ }\bibfield  {title} {\bibinfo {title} {The quench action},\ }\href
  {https://doi.org/10.1088/1742-5468/2016/06/064006} {\bibfield  {journal}
  {\bibinfo  {journal} {Journal of Statistical Mechanics: Theory and
  Experiment}\ }\textbf {\bibinfo {volume} {2016}},\ \bibinfo {pages} {064006}
  (\bibinfo {year} {2016})}\BibitemShut {NoStop}%
\bibitem [{\citenamefont {Piroli}\ \emph {et~al.}(2017)\citenamefont {Piroli},
  \citenamefont {Pozsgay},\ and\ \citenamefont {Vernier}}]{Piroli2017}%
  \BibitemOpen
  \bibfield  {author} {\bibinfo {author} {\bibfnamefont {L.}~\bibnamefont
  {Piroli}}, \bibinfo {author} {\bibfnamefont {B.}~\bibnamefont {Pozsgay}},\
  and\ \bibinfo {author} {\bibfnamefont {E.}~\bibnamefont {Vernier}},\
  }\bibfield  {title} {\bibinfo {title} {What is an integrable quench?},\
  }\href {https://doi.org/https://doi.org/10.1016/j.nuclphysb.2017.10.012}
  {\bibfield  {journal} {\bibinfo  {journal} {Nuclear Physics B}\ }\textbf
  {\bibinfo {volume} {925}},\ \bibinfo {pages} {362} (\bibinfo {year}
  {2017})}\BibitemShut {NoStop}%
\bibitem [{\citenamefont {McGuire}(1964)}]{McGuire1964}%
  \BibitemOpen
  \bibfield  {author} {\bibinfo {author} {\bibfnamefont {J.~B.}\ \bibnamefont
  {McGuire}},\ }\bibfield  {title} {\bibinfo {title} {{Study of exactly soluble
  one‐dimensional N‐body problems}},\ }\href
  {https://doi.org/10.1063/1.1704156} {\bibfield  {journal} {\bibinfo
  {journal} {Journal of Mathematical Physics}\ }\textbf {\bibinfo {volume}
  {5}},\ \bibinfo {pages} {622} (\bibinfo {year} {1964})}\BibitemShut {NoStop}%
\bibitem [{\citenamefont {Piroli}\ \emph {et~al.}(2016)\citenamefont {Piroli},
  \citenamefont {Calabrese},\ and\ \citenamefont {Essler}}]{Piroli2015}%
  \BibitemOpen
  \bibfield  {author} {\bibinfo {author} {\bibfnamefont {L.}~\bibnamefont
  {Piroli}}, \bibinfo {author} {\bibfnamefont {P.}~\bibnamefont {Calabrese}},\
  and\ \bibinfo {author} {\bibfnamefont {F.~H.~L.}\ \bibnamefont {Essler}},\
  }\bibfield  {title} {\bibinfo {title} {Multiparticle bound-state formation
  following a quantum quench to the one-dimensional {B}ose gas with attractive
  interactions},\ }\href {https://doi.org/10.1103/PhysRevLett.116.070408}
  {\bibfield  {journal} {\bibinfo  {journal} {Phys. Rev. Lett.}\ }\textbf
  {\bibinfo {volume} {116}},\ \bibinfo {pages} {070408} (\bibinfo {year}
  {2016})}\BibitemShut {NoStop}%
\bibitem [{\citenamefont {Koch}\ \emph {et~al.}(2021)\citenamefont {Koch},
  \citenamefont {Bastianello},\ and\ \citenamefont {Caux}}]{Koch2021}%
  \BibitemOpen
  \bibfield  {author} {\bibinfo {author} {\bibfnamefont {R.}~\bibnamefont
  {Koch}}, \bibinfo {author} {\bibfnamefont {A.}~\bibnamefont {Bastianello}},\
  and\ \bibinfo {author} {\bibfnamefont {J.-S.}\ \bibnamefont {Caux}},\
  }\bibfield  {title} {\bibinfo {title} {Adiabatic formation of bound states in
  the one-dimensional {B}ose gas},\ }\href
  {https://doi.org/10.1103/PhysRevB.103.165121} {\bibfield  {journal} {\bibinfo
   {journal} {Phys. Rev. B}\ }\textbf {\bibinfo {volume} {103}},\ \bibinfo
  {pages} {165121} (\bibinfo {year} {2021})}\BibitemShut {NoStop}%
\bibitem [{\citenamefont {Bastianello}\ \emph {et~al.}(2019)\citenamefont
  {Bastianello}, \citenamefont {Alba},\ and\ \citenamefont
  {Caux}}]{Bastianello2019}%
  \BibitemOpen
  \bibfield  {author} {\bibinfo {author} {\bibfnamefont {A.}~\bibnamefont
  {Bastianello}}, \bibinfo {author} {\bibfnamefont {V.}~\bibnamefont {Alba}},\
  and\ \bibinfo {author} {\bibfnamefont {J.-S.}\ \bibnamefont {Caux}},\
  }\bibfield  {title} {\bibinfo {title} {Generalized hydrodynamics with
  space-time inhomogeneous interactions},\ }\href
  {https://doi.org/10.1103/PhysRevLett.123.130602} {\bibfield  {journal}
  {\bibinfo  {journal} {Phys. Rev. Lett.}\ }\textbf {\bibinfo {volume} {123}},\
  \bibinfo {pages} {130602} (\bibinfo {year} {2019})}\BibitemShut {NoStop}%
\bibitem [{\citenamefont {M\o{}ller}\ and\ \citenamefont
  {Schmiedmayer}(2020)}]{Moller2020}%
  \BibitemOpen
  \bibfield  {author} {\bibinfo {author} {\bibfnamefont {F.~S.}\ \bibnamefont
  {M\o{}ller}}\ and\ \bibinfo {author} {\bibfnamefont {J.}~\bibnamefont
  {Schmiedmayer}},\ }\bibfield  {title} {\bibinfo {title} {{Introducing iFluid:
  a numerical framework for solving hydrodynamical equations in integrable
  models}},\ }\href {https://doi.org/10.21468/SciPostPhys.8.3.041} {\bibfield
  {journal} {\bibinfo  {journal} {SciPost Phys.}\ }\textbf {\bibinfo {volume}
  {8}},\ \bibinfo {pages} {041} (\bibinfo {year} {2020})}\BibitemShut {NoStop}%
\bibitem [{\citenamefont {De~Nardis}\ and\ \citenamefont
  {Panfil}(2018)}]{DeNardis2018}%
  \BibitemOpen
  \bibfield  {author} {\bibinfo {author} {\bibfnamefont {J.}~\bibnamefont
  {De~Nardis}}\ and\ \bibinfo {author} {\bibfnamefont {M.}~\bibnamefont
  {Panfil}},\ }\bibfield  {title} {\bibinfo {title} {Edge singularities and
  quasilong-range order in nonequilibrium steady states},\ }\href
  {https://doi.org/10.1103/PhysRevLett.120.217206} {\bibfield  {journal}
  {\bibinfo  {journal} {Phys. Rev. Lett.}\ }\textbf {\bibinfo {volume} {120}},\
  \bibinfo {pages} {217206} (\bibinfo {year} {2018})}\BibitemShut {NoStop}%
\bibitem [{\citenamefont {Tak\'acs}\ \emph {et~al.}(2025)\citenamefont
  {Tak\'acs}, \citenamefont {Zhang}, \citenamefont {Calabrese}, \citenamefont
  {Dubail}, \citenamefont {Rigol},\ and\ \citenamefont {Scopa}}]{Takacs2025}%
  \BibitemOpen
  \bibfield  {author} {\bibinfo {author} {\bibfnamefont {A.}~\bibnamefont
  {Tak\'acs}}, \bibinfo {author} {\bibfnamefont {Y.}~\bibnamefont {Zhang}},
  \bibinfo {author} {\bibfnamefont {P.}~\bibnamefont {Calabrese}}, \bibinfo
  {author} {\bibfnamefont {J.}~\bibnamefont {Dubail}}, \bibinfo {author}
  {\bibfnamefont {M.}~\bibnamefont {Rigol}},\ and\ \bibinfo {author}
  {\bibfnamefont {S.}~\bibnamefont {Scopa}},\ }\bibfield  {title} {\bibinfo
  {title} {One-body correlations and momentum distributions of trapped
  one-dimensional {B}ose gases at finite temperature},\ }\href
  {https://doi.org/10.1103/PhysRevA.111.033317} {\bibfield  {journal} {\bibinfo
   {journal} {Phys. Rev. A}\ }\textbf {\bibinfo {volume} {111}},\ \bibinfo
  {pages} {033317} (\bibinfo {year} {2025})}\BibitemShut {NoStop}%
\bibitem [{\citenamefont {Xu}\ and\ \citenamefont {Rigol}(2015)}]{Wei2015}%
  \BibitemOpen
  \bibfield  {author} {\bibinfo {author} {\bibfnamefont {W.}~\bibnamefont
  {Xu}}\ and\ \bibinfo {author} {\bibfnamefont {M.}~\bibnamefont {Rigol}},\
  }\bibfield  {title} {\bibinfo {title} {Universal scaling of density and
  momentum distributions in lieb-liniger gases},\ }\href
  {https://doi.org/10.1103/PhysRevA.92.063623} {\bibfield  {journal} {\bibinfo
  {journal} {Phys. Rev. A}\ }\textbf {\bibinfo {volume} {92}},\ \bibinfo
  {pages} {063623} (\bibinfo {year} {2015})}\BibitemShut {NoStop}%
\bibitem [{\citenamefont {Senese}\ and\ \citenamefont
  {Essler}(2025)}]{senese2025}%
  \BibitemOpen
  \bibfield  {author} {\bibinfo {author} {\bibfnamefont {R.}~\bibnamefont
  {Senese}}\ and\ \bibinfo {author} {\bibfnamefont {F.~H.~L.}\ \bibnamefont
  {Essler}},\ }\href {https://arxiv.org/abs/2508.17908} {\bibinfo {title}
  {Finite temperature single-particle {G}reen's function in the
  {L}ieb-{L}iniger model}} (\bibinfo {year} {2025}),\ \Eprint
  {https://arxiv.org/abs/2508.17908} {arXiv:2508.17908} \BibitemShut {NoStop}%
\bibitem [{\citenamefont {Caux}\ \emph {et~al.}(2007)\citenamefont {Caux},
  \citenamefont {Calabrese},\ and\ \citenamefont {Slavnov}}]{Caux2007}%
  \BibitemOpen
  \bibfield  {author} {\bibinfo {author} {\bibfnamefont {J.-S.}\ \bibnamefont
  {Caux}}, \bibinfo {author} {\bibfnamefont {P.}~\bibnamefont {Calabrese}},\
  and\ \bibinfo {author} {\bibfnamefont {N.~A.}\ \bibnamefont {Slavnov}},\
  }\bibfield  {title} {\bibinfo {title} {One-particle dynamical correlations in
  the one-dimensional {B}ose gas},\ }\href
  {https://doi.org/10.1088/1742-5468/2007/01/P01008} {\bibfield  {journal}
  {\bibinfo  {journal} {Journal of Statistical Mechanics: Theory and
  Experiment}\ }\textbf {\bibinfo {volume} {2007}},\ \bibinfo {pages} {P01008}
  (\bibinfo {year} {2007})}\BibitemShut {NoStop}%
\bibitem [{\citenamefont {Alba}\ and\ \citenamefont
  {Calabrese}(2016)}]{Alba2016}%
  \BibitemOpen
  \bibfield  {author} {\bibinfo {author} {\bibfnamefont {V.}~\bibnamefont
  {Alba}}\ and\ \bibinfo {author} {\bibfnamefont {P.}~\bibnamefont
  {Calabrese}},\ }\bibfield  {title} {\bibinfo {title} {The quench action
  approach in finite integrable spin chains},\ }\href
  {https://doi.org/10.1088/1742-5468/2016/04/043105} {\bibfield  {journal}
  {\bibinfo  {journal} {Journal of Statistical Mechanics: Theory and
  Experiment}\ }\textbf {\bibinfo {volume} {2016}},\ \bibinfo {pages} {043105}
  (\bibinfo {year} {2016})}\BibitemShut {NoStop}%
\bibitem [{\citenamefont {Bouchoule}\ \emph {et~al.}(2020)\citenamefont
  {Bouchoule}, \citenamefont {Doyon},\ and\ \citenamefont
  {Dubail}}]{Dubail2020}%
  \BibitemOpen
  \bibfield  {author} {\bibinfo {author} {\bibfnamefont {I.}~\bibnamefont
  {Bouchoule}}, \bibinfo {author} {\bibfnamefont {B.}~\bibnamefont {Doyon}},\
  and\ \bibinfo {author} {\bibfnamefont {J.}~\bibnamefont {Dubail}},\
  }\bibfield  {title} {\bibinfo {title} {{The effect of atom losses on the
  distribution of rapidities in the one-dimensional {B}ose gas}},\ }\href
  {https://doi.org/10.21468/SciPostPhys.9.4.044} {\bibfield  {journal}
  {\bibinfo  {journal} {SciPost Phys.}\ }\textbf {\bibinfo {volume} {9}},\
  \bibinfo {pages} {044} (\bibinfo {year} {2020})}\BibitemShut {NoStop}%
\bibitem [{\citenamefont {Zhang}\ \emph {et~al.}(2024)\citenamefont {Zhang},
  \citenamefont {Yu}, \citenamefont {Chen}, \citenamefont {Cheng},\ and\
  \citenamefont {Guan}}]{Zhang2024}%
  \BibitemOpen
  \bibfield  {author} {\bibinfo {author} {\bibfnamefont {Z.-H.}\ \bibnamefont
  {Zhang}}, \bibinfo {author} {\bibfnamefont {Y.-C.}\ \bibnamefont {Yu}},
  \bibinfo {author} {\bibfnamefont {Y.-Y.}\ \bibnamefont {Chen}}, \bibinfo
  {author} {\bibfnamefont {S.}~\bibnamefont {Cheng}},\ and\ \bibinfo {author}
  {\bibfnamefont {X.-W.}\ \bibnamefont {Guan}},\ }\bibfield  {title} {\bibinfo
  {title} {Monte {C}arlo {B}ethe-ansatz approach for the study of the
  {L}ieb-{L}iniger model},\ }\href
  {https://doi.org/10.1103/PhysRevA.109.033320} {\bibfield  {journal} {\bibinfo
   {journal} {Phys. Rev. A}\ }\textbf {\bibinfo {volume} {109}},\ \bibinfo
  {pages} {033320} (\bibinfo {year} {2024})}\BibitemShut {NoStop}%
\bibitem [{\citenamefont {De~Nardis}\ \emph {et~al.}(2018)\citenamefont
  {De~Nardis}, \citenamefont {Bernard},\ and\ \citenamefont
  {Doyon}}]{Nardis2018}%
  \BibitemOpen
  \bibfield  {author} {\bibinfo {author} {\bibfnamefont {J.}~\bibnamefont
  {De~Nardis}}, \bibinfo {author} {\bibfnamefont {D.}~\bibnamefont {Bernard}},\
  and\ \bibinfo {author} {\bibfnamefont {B.}~\bibnamefont {Doyon}},\ }\bibfield
   {title} {\bibinfo {title} {Hydrodynamic diffusion in integrable systems},\
  }\href {https://doi.org/10.1103/PhysRevLett.121.160603} {\bibfield  {journal}
  {\bibinfo  {journal} {Phys. Rev. Lett.}\ }\textbf {\bibinfo {volume} {121}},\
  \bibinfo {pages} {160603} (\bibinfo {year} {2018})}\BibitemShut {NoStop}%
\bibitem [{\citenamefont {Bastianello}\ \emph {et~al.}(2020)\citenamefont
  {Bastianello}, \citenamefont {De~Luca}, \citenamefont {Doyon},\ and\
  \citenamefont {De~Nardis}}]{Bastianello2020}%
  \BibitemOpen
  \bibfield  {author} {\bibinfo {author} {\bibfnamefont {A.}~\bibnamefont
  {Bastianello}}, \bibinfo {author} {\bibfnamefont {A.}~\bibnamefont
  {De~Luca}}, \bibinfo {author} {\bibfnamefont {B.}~\bibnamefont {Doyon}},\
  and\ \bibinfo {author} {\bibfnamefont {J.}~\bibnamefont {De~Nardis}},\
  }\bibfield  {title} {\bibinfo {title} {Thermalization of a trapped
  one-dimensional {B}ose gas via diffusion},\ }\href
  {https://doi.org/10.1103/PhysRevLett.125.240604} {\bibfield  {journal}
  {\bibinfo  {journal} {Phys. Rev. Lett.}\ }\textbf {\bibinfo {volume} {125}},\
  \bibinfo {pages} {240604} (\bibinfo {year} {2020})}\BibitemShut {NoStop}%
\bibitem [{\citenamefont {Durnin}\ \emph {et~al.}(2021)\citenamefont {Durnin},
  \citenamefont {De~Luca}, \citenamefont {De~Nardis},\ and\ \citenamefont
  {Doyon}}]{Durnin2021}%
  \BibitemOpen
  \bibfield  {author} {\bibinfo {author} {\bibfnamefont {J.}~\bibnamefont
  {Durnin}}, \bibinfo {author} {\bibfnamefont {A.}~\bibnamefont {De~Luca}},
  \bibinfo {author} {\bibfnamefont {J.}~\bibnamefont {De~Nardis}},\ and\
  \bibinfo {author} {\bibfnamefont {B.}~\bibnamefont {Doyon}},\ }\bibfield
  {title} {\bibinfo {title} {Diffusive hydrodynamics of inhomogenous
  hamiltonians},\ }\href {https://doi.org/10.1088/1751-8121/ac2c57} {\bibfield
  {journal} {\bibinfo  {journal} {Journal of Physics A: Mathematical and
  Theoretical}\ }\textbf {\bibinfo {volume} {54}},\ \bibinfo {pages} {494001}
  (\bibinfo {year} {2021})}\BibitemShut {NoStop}%
\bibitem [{\citenamefont {H\"ubner}\ \emph {et~al.}(2025)\citenamefont
  {H\"ubner}, \citenamefont {Biagetti}, \citenamefont {De~Nardis},\ and\
  \citenamefont {Doyon}}]{Hubner2025}%
  \BibitemOpen
  \bibfield  {author} {\bibinfo {author} {\bibfnamefont {F.}~\bibnamefont
  {H\"ubner}}, \bibinfo {author} {\bibfnamefont {L.}~\bibnamefont {Biagetti}},
  \bibinfo {author} {\bibfnamefont {J.}~\bibnamefont {De~Nardis}},\ and\
  \bibinfo {author} {\bibfnamefont {B.}~\bibnamefont {Doyon}},\ }\bibfield
  {title} {\bibinfo {title} {Diffusive hydrodynamics from long-range
  correlations},\ }\href {https://doi.org/10.1103/PhysRevLett.134.187101}
  {\bibfield  {journal} {\bibinfo  {journal} {Phys. Rev. Lett.}\ }\textbf
  {\bibinfo {volume} {134}},\ \bibinfo {pages} {187101} (\bibinfo {year}
  {2025})}\BibitemShut {NoStop}%
\bibitem [{\citenamefont {Caux}\ and\ \citenamefont {Essler}(2013)}]{Caux2013}%
  \BibitemOpen
  \bibfield  {author} {\bibinfo {author} {\bibfnamefont {J.-S.}\ \bibnamefont
  {Caux}}\ and\ \bibinfo {author} {\bibfnamefont {F.~H.~L.}\ \bibnamefont
  {Essler}},\ }\bibfield  {title} {\bibinfo {title} {Time evolution of local
  observables after quenching to an integrable model},\ }\href
  {https://doi.org/10.1103/PhysRevLett.110.257203} {\bibfield  {journal}
  {\bibinfo  {journal} {Phys. Rev. Lett.}\ }\textbf {\bibinfo {volume} {110}},\
  \bibinfo {pages} {257203} (\bibinfo {year} {2013})}\BibitemShut {NoStop}%
\bibitem [{\citenamefont {Essler}\ and\ \citenamefont
  {de~Klerk}(2024)}]{Essler2024}%
  \BibitemOpen
  \bibfield  {author} {\bibinfo {author} {\bibfnamefont {F.~H.~L.}\
  \bibnamefont {Essler}}\ and\ \bibinfo {author} {\bibfnamefont {A.~J. J.~M.}\
  \bibnamefont {de~Klerk}},\ }\bibfield  {title} {\bibinfo {title} {Statistics
  of matrix elements of local operators in integrable models},\ }\href
  {https://doi.org/10.1103/PhysRevX.14.031048} {\bibfield  {journal} {\bibinfo
  {journal} {Phys. Rev. X}\ }\textbf {\bibinfo {volume} {14}},\ \bibinfo
  {pages} {031048} (\bibinfo {year} {2024})}\BibitemShut {NoStop}%
\bibitem [{\citenamefont {Koch}\ \emph {et~al.}(2022)\citenamefont {Koch},
  \citenamefont {Caux},\ and\ \citenamefont {Bastianello}}]{Koch2022}%
  \BibitemOpen
  \bibfield  {author} {\bibinfo {author} {\bibfnamefont {R.}~\bibnamefont
  {Koch}}, \bibinfo {author} {\bibfnamefont {J.-S.}\ \bibnamefont {Caux}},\
  and\ \bibinfo {author} {\bibfnamefont {A.}~\bibnamefont {Bastianello}},\
  }\bibfield  {title} {\bibinfo {title} {Generalized hydrodynamics of the
  attractive non-linear {S}chr\"odinger equation},\ }\href
  {https://doi.org/10.1088/1751-8121/ac53c3} {\bibfield  {journal} {\bibinfo
  {journal} {Journal of Physics A: Mathematical and Theoretical}\ }\textbf
  {\bibinfo {volume} {55}},\ \bibinfo {pages} {134001} (\bibinfo {year}
  {2022})}\BibitemShut {NoStop}%
\bibitem [{\citenamefont {Chen}\ and\ \citenamefont {Cui}(2023)}]{Chen2023}%
  \BibitemOpen
  \bibfield  {author} {\bibinfo {author} {\bibfnamefont {Y.}~\bibnamefont
  {Chen}}\ and\ \bibinfo {author} {\bibfnamefont {X.}~\bibnamefont {Cui}},\
  }\bibfield  {title} {\bibinfo {title} {{Ultrastable super Tonks-Girardeau
  gases under weak dipolar interactions}},\ }\href
  {https://doi.org/10.1103/PhysRevLett.131.203002} {\bibfield  {journal}
  {\bibinfo  {journal} {Phys. Rev. Lett.}\ }\textbf {\bibinfo {volume} {131}},\
  \bibinfo {pages} {203002} (\bibinfo {year} {2023})}\BibitemShut {NoStop}%
\bibitem [{\citenamefont {Gerton}\ \emph {et~al.}(2000)\citenamefont {Gerton},
  \citenamefont {Strekalov}, \citenamefont {Prodan},\ and\ \citenamefont
  {Hulet}}]{Gerton2000}%
  \BibitemOpen
  \bibfield  {author} {\bibinfo {author} {\bibfnamefont {J.~M.}\ \bibnamefont
  {Gerton}}, \bibinfo {author} {\bibfnamefont {D.}~\bibnamefont {Strekalov}},
  \bibinfo {author} {\bibfnamefont {I.}~\bibnamefont {Prodan}},\ and\ \bibinfo
  {author} {\bibfnamefont {R.~G.}\ \bibnamefont {Hulet}},\ }\bibfield  {title}
  {\bibinfo {title} {{Direct observation of growth and collapse of a
  {B}ose--Einstein condensate with attractive interactions}},\ }\href
  {https://doi.org/10.1038/35047030} {\bibfield  {journal} {\bibinfo  {journal}
  {Nature}\ }\textbf {\bibinfo {volume} {408}},\ \bibinfo {pages} {692}
  (\bibinfo {year} {2000})}\BibitemShut {NoStop}%
\bibitem [{\citenamefont {Donley}\ \emph {et~al.}(2001)\citenamefont {Donley},
  \citenamefont {Claussen}, \citenamefont {Cornish}, \citenamefont {Roberts},
  \citenamefont {Cornell},\ and\ \citenamefont {Wieman}}]{Donley2001}%
  \BibitemOpen
  \bibfield  {author} {\bibinfo {author} {\bibfnamefont {E.~A.}\ \bibnamefont
  {Donley}}, \bibinfo {author} {\bibfnamefont {N.~R.}\ \bibnamefont
  {Claussen}}, \bibinfo {author} {\bibfnamefont {S.~L.}\ \bibnamefont
  {Cornish}}, \bibinfo {author} {\bibfnamefont {J.~L.}\ \bibnamefont
  {Roberts}}, \bibinfo {author} {\bibfnamefont {E.~A.}\ \bibnamefont
  {Cornell}},\ and\ \bibinfo {author} {\bibfnamefont {C.~E.}\ \bibnamefont
  {Wieman}},\ }\bibfield  {title} {\bibinfo {title} {{Dynamics of collapsing
  and exploding {B}ose--Einstein condensates}},\ }\href
  {https://doi.org/10.1038/35085500} {\bibfield  {journal} {\bibinfo  {journal}
  {Nature}\ }\textbf {\bibinfo {volume} {412}},\ \bibinfo {pages} {295}
  (\bibinfo {year} {2001})}\BibitemShut {NoStop}%
\bibitem [{\citenamefont {Strecker}\ \emph {et~al.}(2002)\citenamefont
  {Strecker}, \citenamefont {Partridge}, \citenamefont {Truscott},\ and\
  \citenamefont {Hulet}}]{Strecker2002}%
  \BibitemOpen
  \bibfield  {author} {\bibinfo {author} {\bibfnamefont {K.~E.}\ \bibnamefont
  {Strecker}}, \bibinfo {author} {\bibfnamefont {G.~B.}\ \bibnamefont
  {Partridge}}, \bibinfo {author} {\bibfnamefont {A.~G.}\ \bibnamefont
  {Truscott}},\ and\ \bibinfo {author} {\bibfnamefont {R.~G.}\ \bibnamefont
  {Hulet}},\ }\bibfield  {title} {\bibinfo {title} {Formation and propagation
  of matter-wave soliton trains},\ }\href {https://doi.org/10.1038/nature747}
  {\bibfield  {journal} {\bibinfo  {journal} {Nature}\ }\textbf {\bibinfo
  {volume} {417}},\ \bibinfo {pages} {150} (\bibinfo {year}
  {2002})}\BibitemShut {NoStop}%
\bibitem [{\citenamefont {Panfil}\ \emph {et~al.}(2023)\citenamefont {Panfil},
  \citenamefont {Gopalakrishnan},\ and\ \citenamefont {Konik}}]{Panfil2023}%
  \BibitemOpen
  \bibfield  {author} {\bibinfo {author} {\bibfnamefont {M.}~\bibnamefont
  {Panfil}}, \bibinfo {author} {\bibfnamefont {S.}~\bibnamefont
  {Gopalakrishnan}},\ and\ \bibinfo {author} {\bibfnamefont {R.~M.}\
  \bibnamefont {Konik}},\ }\bibfield  {title} {\bibinfo {title} {Thermalization
  of interacting quasi-one-dimensional systems},\ }\href
  {https://doi.org/10.1103/PhysRevLett.130.030401} {\bibfield  {journal}
  {\bibinfo  {journal} {Phys. Rev. Lett.}\ }\textbf {\bibinfo {volume} {130}},\
  \bibinfo {pages} {030401} (\bibinfo {year} {2023})}\BibitemShut {NoStop}%
\bibitem [{\citenamefont {Biagetti}\ \emph {et~al.}(2025)\citenamefont
  {Biagetti}, \citenamefont {Lebek}, \citenamefont {Panfil},\ and\
  \citenamefont {Nardis}}]{biagetti2025}%
  \BibitemOpen
  \bibfield  {author} {\bibinfo {author} {\bibfnamefont {L.}~\bibnamefont
  {Biagetti}}, \bibinfo {author} {\bibfnamefont {M.}~\bibnamefont {Lebek}},
  \bibinfo {author} {\bibfnamefont {M.}~\bibnamefont {Panfil}},\ and\ \bibinfo
  {author} {\bibfnamefont {J.~D.}\ \bibnamefont {Nardis}},\ }\href
  {https://arxiv.org/abs/2408.00593} {\bibinfo {title} {Generalised {BBGKY}
  hierarchy for near-integrable dynamics}} (\bibinfo {year} {2025}),\ \Eprint
  {https://arxiv.org/abs/2408.00593} {arXiv:2408.00593} \BibitemShut {NoStop}%
\bibitem [{Zen()}]{Zenodo}%
  \BibitemOpen
  \href@noop {} {}\bibinfo {howpublished} {{Research data are freely available
  from Zenodo at doi.org/10.5281/zenodo.18702729}}\BibitemShut {NoStop}%
\bibitem [{\citenamefont {Doyon}(2020)}]{Doyon2020}%
  \BibitemOpen
  \bibfield  {author} {\bibinfo {author} {\bibfnamefont {B.}~\bibnamefont
  {Doyon}},\ }\bibfield  {title} {\bibinfo {title} {{Lecture notes on
  Generalised Hydrodynamics}},\ }\href
  {https://doi.org/10.21468/SciPostPhysLectNotes.18} {\bibfield  {journal}
  {\bibinfo  {journal} {SciPost Phys. Lect. Notes}\ ,\ \bibinfo {pages} {18}}
  (\bibinfo {year} {2020})}\BibitemShut {NoStop}%
\bibitem [{\citenamefont {Kheruntsyan}\ \emph {et~al.}(2003)\citenamefont
  {Kheruntsyan}, \citenamefont {Gangardt}, \citenamefont {Drummond},\ and\
  \citenamefont {Shlyapnikov}}]{Gangardt2003}%
  \BibitemOpen
  \bibfield  {author} {\bibinfo {author} {\bibfnamefont {K.~V.}\ \bibnamefont
  {Kheruntsyan}}, \bibinfo {author} {\bibfnamefont {D.~M.}\ \bibnamefont
  {Gangardt}}, \bibinfo {author} {\bibfnamefont {P.~D.}\ \bibnamefont
  {Drummond}},\ and\ \bibinfo {author} {\bibfnamefont {G.~V.}\ \bibnamefont
  {Shlyapnikov}},\ }\bibfield  {title} {\bibinfo {title} {Pair correlations in
  a finite-temperature 1d {B}ose gas},\ }\href
  {https://doi.org/10.1103/PhysRevLett.91.040403} {\bibfield  {journal}
  {\bibinfo  {journal} {Phys. Rev. Lett.}\ }\textbf {\bibinfo {volume} {91}},\
  \bibinfo {pages} {040403} (\bibinfo {year} {2003})}\BibitemShut {NoStop}%
\bibitem [{\citenamefont {Doyon}\ and\ \citenamefont
  {Yoshimura}(2017)}]{Doyon2017}%
  \BibitemOpen
  \bibfield  {author} {\bibinfo {author} {\bibfnamefont {B.}~\bibnamefont
  {Doyon}}\ and\ \bibinfo {author} {\bibfnamefont {T.}~\bibnamefont
  {Yoshimura}},\ }\bibfield  {title} {\bibinfo {title} {{A note on generalized
  hydrodynamics: inhomogeneous fields and other concepts}},\ }\href
  {https://doi.org/10.21468/SciPostPhys.2.2.014} {\bibfield  {journal}
  {\bibinfo  {journal} {SciPost Phys.}\ }\textbf {\bibinfo {volume} {2}},\
  \bibinfo {pages} {014} (\bibinfo {year} {2017})}\BibitemShut {NoStop}%
\end{thebibliography}%

\clearpage
\onecolumngrid
\newpage

\setcounter{equation}{0}  
\setcounter{figure}{0}
% reset equation counter
\setcounter{page}{1}
\setcounter{section}{0}    % reset section counter
\renewcommand\thesection{\arabic{section}}    % puts letters as section numbering
\renewcommand\thesubsection{\arabic{subsection}}    % puts letters as section numbering
\renewcommand{\thetable}{S\arabic{table}}
\renewcommand{\theequation}{S\arabic{equation}}
\renewcommand{\thefigure}{S\arabic{figure}}
\setcounter{secnumdepth}{2}  % if the subsections need to be numbered

\begin{center}
{\Large Supplementary Material\\
\titleinfo
} 
\end{center}
\bigskip
\bigskip

The Supplementary Material provides the technical details supporting our results. Section~\ref{sec_TBAGHD} overviews the thermodynamics of integrable models and their hydrodynamics (GHD), while Sec.~\ref{sec_MC} discusses the numerical method we used to compute the momentum distribution and $g^{(1)}(x)$.

\section{Thermodynamic Bethe ansatz and Generalized Hydrodynamics }
\label{sec_TBAGHD}

In this Section we provide a short overview of the thermodynamics of integrable models, usually referred to as thermodynamic Bethe ansatz (TBA)~\cite{takahashi2005thermodynamics}, and Generalized Hydrodynamics (GHD). Here, we give a short account of these frameworks: for a more in-depth discussion, the interested reader can refer to reviews~\cite{Bastianello2022,Doyon2024} and lecture notes~\cite{Doyon2020}.

\subsection{From microscopic to thermodynamics}
\label{sec_tba}
Integrable models feature infinitely-many extensive conserved charges $\hat{Q}_j=\int \dd x\, \hat{{\rm q}}_j(x)$, where $\hat{{\rm q}}_j(x)$ is a local observable. In this ensemble of charges, one can find the number or particles $\hat{{\rm q}}_j(x)\to \hat{\psi}^\dagger(x)\hat{\psi}(x)$, the momentum $\hat{{\rm q}}_j(x)\to \frac{\hbar}{2}(i\hat{\psi}^\dagger(x)\partial_x\hat{\psi}(x)+\text{h.c.})$, the Hamiltonian $\hat{{\rm q}}_j(x)\to \frac{\hbar^2}{2m}\partial_x\hat{\psi}^\dagger\partial_x\hat{\psi}+\frac{\g}{2}\hat{\psi}^\dagger\hat{\psi}^\dagger\hat{\psi}\hat{\psi}$, and infinitely many others~\cite{takahashi2005thermodynamics}.
These extensive charges are in involution, and simultaneous eigenstates can be found in the form
\be\label{eq_suppstate}
|\{\lambda_i\}_{i=1}^N\rangle\propto \sum_{\mathcal{P}}A(\mathcal{P})\prod_{j=1}^N e^{i\lambda_{\mathcal{P}_j}x_j}\, ,
\ee
where above one sums over all the permutations $\mathcal{P}$ of the rapidity's set $\{\lambda_i\}_{i=1}^N$, and the coefficients obey $A(\Pi_{j,j+1}\mathcal{P})=-\exp\left[-i \Theta(\lambda_{\mathcal{P}_j}-\lambda_{\mathcal{P}_{j+1}})\right]A(\mathcal{P})$ with $\Pi_{j,j+1}$ being the permutation exchanging the elements in positions $j$ and $j+1$.
Similarly to what happens in non-interacting systems, the conserved charges act additively on the rapidities $\hat{Q}_j|\{ \lambda_i\}_{i=1}^N\rangle=\left(\sum_{i=1}^N q_j(\lambda_i)\right)|\{\lambda_i\}_{i=1}^N\rangle$, where the functions $q_j(\lambda)$ are called charge eigenvalues. For the number of particles $q_j(\lambda)\to 1$, for the momentum $q_j(\lambda)\to p(\lambda)=\hbar \lambda$, and for the energy $q_j(\lambda)\to \epsilon(\lambda)=\frac{\hbar^2 \lambda^2}{2m}$. On a finite system of size $L$, periodic boundary conditions result in non-linear equations to quantize the allowed rapidities, known as Bethe equations~\cite{Lieb1963}
\be\label{Seq_bethe}
\frac{I_j}{L}=\frac{\lambda_j}{2\pi}+\frac{1}{2\pi L}\sum_{k\ne j} \Theta(\lambda_j-\lambda_k)  \, ,
\ee
where $\{I_j\}_{j=1}^N$ are integers for odd number of particles, and half integers for even number of particles. Any set of $\{I_j\}_{j=1}^N$ with all distinct elements corresponds to an eigenstate \eqref{eq_suppstate}. In the repulsive regime of LL $(\g>0)$, the Bethe equations admit only solutions with real rapidities. In the attractive regime, complex rapidities are possible and, in the thermodynamic limit, they arrange in regular patterns where ``strings" of rapidities share the same real part and are regularly shifted in the imaginary axis. These solutions are called Bethe strings~\cite{takahashi2005thermodynamics} and are associated with bound states. Hereafter, we assume no bound states are excited, so only real rapidities are possible: a short detour accounting for bound states will be provided in Section~\ref{sec_ghd} when discussing the irreversibility of the cycle. The Bethe equations are highly non-linear and difficult to solve, but one can largely simplify the problem in the thermodynamic limit introducing the root density $\rho(\lambda)$ as $\frac{I_j}{L}=\frac{\lambda_j}{2\pi}+\int\frac{\dd\lambda'}{2\pi} \Theta(\lambda_j-\lambda')\rho(\lambda')+\mathcal{O}(L^{-2})$. As a next step, one looks at these equations as a state-dependent change of variables $\lambda\to z(\lambda)=\frac{\lambda}{2\pi}+\int\frac{\dd\lambda'}{2\pi} \Theta(\lambda-\lambda')\rho(\lambda')$. The total root density is then introduced as the Jacobian of this change of variables $\rho^t(\lambda)\equiv \tfrac{\dd z(\lambda)}{\dd\lambda}$.
We can now discuss thermodynamics and, more generally, GGEs. The grand canonical partition function on a Gibbs ensemble is $\mathcal{Z}_{\text{Gibbs}}=\text{Tr}\left[e^{-\beta (\hat{H}-\mu \hat{N})}\right]$, with $\hat{N}$ being the number of particles, while $\beta=\frac{1}{K_\text{B} T}$ and $\mu$ is the chemical potential. On more general GGEs, the partition function is~\cite{Rigol2007} $\mathcal{Z}_\text{GGE}=\text{Tr}\left[e^{-\sum_k \beta_k \hat{Q}_k}\right]$. The trace over the GGE density matrix can be written on the basis of the quantum numbers $\{I_j\}_{j=1}^N$ and then, in the thermodynamic limit, converted into a path integral over the possible macroscopic configurations in the rapidity space $\rho(\lambda)$~\cite{takahashi2005thermodynamics}
\be\label{eq_ZGGE}
\mathcal{Z}_{\text{GGE}}=\sum_{\{I_j\}_{j=1}^N} e^{-\sum_k \sum_{j=1}^N\beta_k q_k(\lambda_j)}\simeq \int \mathcal{D}\rho\, e^{\mathcal{S}[\rho]-L\int \dd\lambda\, \left[\sum_k\beta_k q_k(\lambda)\right]\rho(\lambda)}\, ,
\ee
where the entropy 
\be\label{Seq_entropy}
\mathcal{S}[\rho]=L\int \dd\lambda\, \rho^t(\lambda)\left[-\vartheta(\lambda) \log\vartheta(\lambda)-(1-\vartheta(\lambda))\log(1-\vartheta(\lambda))\right]\, ,
\ee
accounts for the microstate degeneracy associated to a macrostate $\rho(\lambda)$, and we defined the occupancy $\vartheta(\lambda)=\rho(\lambda)/\rho^t(\lambda)$. In the thermodynamic limit $L\to +\infty$, the saddle point approximation of the path integral becomes exact, resulting in an integral equation uniquely determining the root density $\rho(\lambda)$ associated to a set of generalized temperatures $\{\beta_k\}_k$
\be\label{eq_tba}
\varepsilon(\lambda)=\sum_k \beta_k q_k(\lambda)+\int \frac{\dd\lambda'}{2\pi}\varphi(\lambda-\lambda')\log(1+e^{-\varepsilon(\lambda')})\, ,
\ee
where we parametrized the occupancy as $\vartheta(\lambda)=\big(1+e^{\varepsilon(\lambda)}\big)^{-1}$ and we used $\varphi(\lambda)=\partial_\lambda \Theta(\lambda)$.
If the form of a GGE in terms of the generalized inverse temperature is known (hence, $\sum_k \beta_k q_k(\lambda)$ is determined), the above equation ultimately allows to determine the GGE's occupancy $\vartheta(\lambda)$ and root density $\rho(\lambda)$. The reverse is also true: knowing the occupancy of a GGE, through Eq.~\eqref{eq_tba} one can fix the function $\sum_k \beta_k q_k(\lambda)$ and ultimately characterize the GGE microscopically through Eq.~\eqref{eq_ZGGE}. We will take advantage of this fact to sample the momentum distribution in Section~\ref{sec_MC}.
Let us now focus on Gibbs ensembles $\sum_k \beta_k q_k(\lambda)\to \beta(\epsilon(\lambda)-\mu)$ and more precisely on the ground state sending the temperature to zero $\beta\to +\infty$. In this case, Eq.~\eqref{eq_tba} becomes singular and $\varepsilon(\lambda)$ diverges to negative values within an interval $\lambda\in (-\lambda_F,\lambda_F)$, whereas it diverges to positive values out of it. As a result, the occupancy becomes a Fermi sea with unitary filling $\vartheta(\lambda)=\chi(\lambda,\lambda_F)$, where the value of the Fermi rapidity $\lambda_F$ is implicitly determined fixing the particle's density $n=\int \dd\lambda\, \rho(\lambda)$.
Before discussing hydrodynamics, we report the expressions to compute the energies shown in Fig.~2.
Since the total energy is a conserved charge, its expression readily follows
\be\label{Seq_Etot}
E_\text{tot}=\Bigg\langle \frac{\hbar^2}{2m}\partial_x\hat{\psi}^\dagger\partial_x\hat{\psi}+\frac{\g}{2}\hat{\psi}^\dagger\hat{\psi}^\dagger\hat{\psi}\hat{\psi}\Bigg\rangle=\int \dd\lambda \,\epsilon(\lambda) \rho(\lambda)\, .
\ee
To access the kinetic energy $E_\text{kin}=\Bigg\langle \frac{\hbar^2}{2m}\partial_x\hat{\psi}^\dagger\partial_x\hat{\psi}\Bigg\rangle$, which is of utmost interest for cold atom experiments, one uses $E_\text{tot}-E_\text{kin}=\frac{\g}{2}\langle \hat{\psi}^\dagger\hat{\psi}^\dagger\hat{\psi}\hat{\psi}\rangle$, where $\langle\hat{\psi}^\dagger\hat{\psi}^\dagger\hat{\psi}\hat{\psi}\rangle$ can be computed through the Hellmann-Feynman theorem~\cite{Gangardt2003}
\be\label{Seq_g2}
\langle\hat{\psi}^\dagger\hat{\psi}^\dagger\hat{\psi}\hat{\psi}\rangle=\int \frac{\dd\lambda}{2\pi}(\partial_\lambda \epsilon)^\dr\vartheta(\lambda) \int \dd\lambda' \partial_{\g}\Theta(\lambda-\lambda')\rho(\lambda')\, ,
\ee
where for an arbitrary function $\tau(\lambda)$ one defines the dressing $\tau(\lambda)\to \tau^\dr(\lambda)$ as the solution of the integral equation $\tau^\dr(\lambda)=\tau(\lambda)-\int \frac{\dd\lambda'}{2\pi}\varphi(\lambda-\lambda')\vartheta(\lambda')\tau^\dr(\lambda')$.

\subsection{Hydrodynamics in integrable models}
\label{sec_ghd}

The hydrodynamics of integrable models is framed within Generalized Hydrodynamics (GHD). Initially proposed to describe inhomogeneous states evolved with exactly integrable Hamiltonians~\cite{Alba2016,Bertini2016}, it quickly grew into a powerful framework capable of handling integrability breaking~\cite{Bastianello2022} and ultimately quantitatively describing experiments~\cite{Schemmer2019,Malvania2021,Schuttelkopf2024,dubois2024,horvath2025,Cataldini2022,Moller2021,Yang2024Phantom}.
GHD assumes a separation of scale, approximating the local properties of the system as well-described by a local GGE $\rho(\lambda)\to \rho_{t,x}(\lambda)$: on a large scale, the local and instantaneous root density evolves accordingly to certain kinetic equations. At the Euler scale, i.e. by retaining only up to first derivatives in the hydrodynamic expansion, it states
\be\label{Seq_GHD_rho}
\partial_t \rho_{t,x}(\lambda)+\partial_x[v^\text{eff}(\lambda)\rho_{t,x}(\lambda)]+\partial_\lambda[a^\text{eff}(\lambda) \rho_{t,x}(\lambda)]=0\, ,
\ee
that can be equivalently rewritten in terms of the occupancy as
\be\label{Seq_GHD_fill}
\partial_t \vartheta_{t,x}(\lambda)+v^\text{eff}(\lambda)\partial_x\vartheta_{t,x}(\lambda)+a^\text{eff}(\lambda) \partial_\lambda\vartheta_{t,x}(\lambda)=0\, .
\ee

The effective velocity $v^\eff(\lambda)$ and acceleration $a^\eff(\lambda)$ are renormalized by the effect of interactions and depend on the local state $\rho_{t,x}(\lambda)$, making the GHD equations highly non-linear. The effective acceleration is caused by weak integrability breaking, for example, by adding a smooth potential to the LL Hamiltonian $\hat{H}\to\hat{H}+\int \dd x\, V(x)\hat{\psi}^\dagger\hat{\psi}$ or by considering a time-dependent coupling.
In this case, the effective velocity~\cite{Alvaredo2016,Bertini2016} and acceleration~\cite{Doyon2017,Bastianello2019} have the following form
\be
v^\eff(\lambda)=\frac{(\partial_\lambda \epsilon)^\dr}{(\partial_\lambda p)^\dr}\, , \hspace{2pc}a^\eff(\lambda)=\frac{(f(\lambda))^\dr}{(\partial_\lambda p)^\dr}\, ,
\ee
where the bare force term is~\cite{Bastianello2019,Moller2020}
\be
f(\lambda)= -\partial_x V+\partial_t \g\int \dd\lambda' \partial_{\g}\Theta(\lambda-\lambda')\rho(\lambda')\, .
\ee

Although we gave the full inhomogeneous form of the GHD equation \eqref{Seq_GHD_rho} \eqref{Seq_GHD_fill} for completeness, in this work for the sake of simplicity we focus on the homogeneous case, hence we drop spatial dependencies. In realistic experiments~\cite{ExpHolo}, the force term due to the inhomogeneity must be accounted for. In this case, one resorts to a numerical solution of the GHD equations based on the implicit solution within the method of characteristics~\cite{Bastianello2019,Moller2020}
\be\label{eq_characteristics}
\vartheta_{t,x(t)}(\lambda(t))=\vartheta_{t_0,x}(\lambda)
\ee
where the evolving coordinates $x(t)$ and $\lambda(t)$ satisfies
\be
\frac{\dd x(t)}{\dd t}=v^\eff(\lambda(t))\Big|_{x=x(t)} \hspace{2pc} \frac{\dd \lambda(t)}{\dd t}=a^\eff(\lambda(t))\Big|_{x=x(t)} 
\ee
with initial conditions $x(t_0)=x$ and $\lambda(t_0)=\lambda$. This implicit solution is very convenient for numerical methods~\cite{Bastianello2019,Moller2020}, and it describes an inhomogeneous translation of the filling fraction in the phase space. It immediately follows that $\max\vartheta_{t,x}(\lambda)$ remains constant during the evolution with Eq. \eqref{eq_characteristics}. Hence, the maximum occupancy $\vartheta$ does not change within the repulsive or attractive branches, but it can jump when crossing between the two regimes.

The effect of the time-dependent coupling can be propagated with GHD equations until the TG-sTG transition, then continued smoothly in the attractive case where bound states are not excited, and eventually reach the non-interacting point. In crossing the non-interacting point from the attractive to the repulsive side $\g=0^{-}\to \g=0^+$, one must ensure the continuity of $\rho(\lambda)$ resulting in a discontinuity of the occupancy $\vartheta(\lambda)$ as discussed in the main text, and a sudden entropy increase.
The GHD equations \eqref{Seq_GHD_rho} \eqref{Seq_GHD_fill} conserve the entropy \eqref{Seq_entropy} $\partial_t \mathcal{S}=0$~\cite{Bastianello2019}, but sudden jumps can appear whenever crossing the non-interacting points. To compute the staircase value of the entropy it is most convenient to focus on the TG regime, where the scattering kernel becomes trivial $\lim_{\g\to\infty}\varphi(\lambda)=0$: here, the result announced in the main text Eq.~(4) is readily recovered.

\subsubsection{Crossing $\g=0$ from repulsive to attractive: bound state formation and reversibility-breakdown}

\begin{figure}[h!]
\includegraphics[width=0.99\columnwidth]{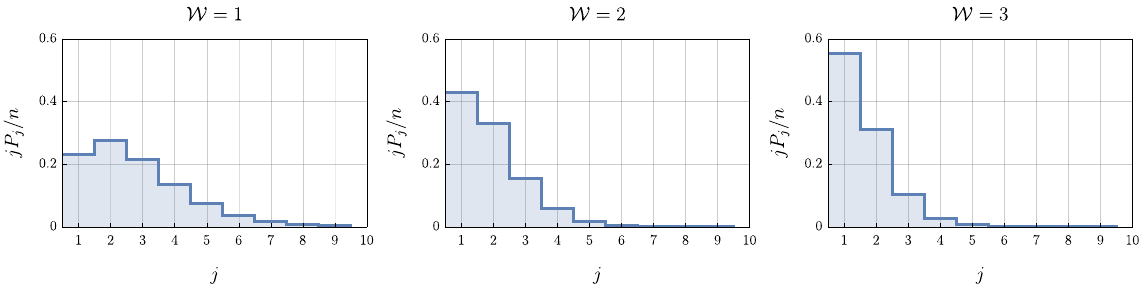}
\caption{\textbf{Bound state production in the reversed cycle.---} We show the bound state participation, normalized on the particle's density $n$, obtained by reversing the cycle and crossing $\g=0^+\to \g=0^-$ after having first reached the cycle's winding $\wind$.
As $\wind$ is increased, the formation of large bound states is more unlikely.
}\label{Fig_S1}
\end{figure}

In principle, the attractive regime features bound states, called Bethe strings, of an arbitrary number of particles. According to GHD, these bound states are not excited when crossing the TG-sTG transition, nor during the slow evolution through the attractive regime in the direct cycle, hence we neglected them so far. However, reversing the cycle, bound states are created crossing $\g=0$ from the repulsive $\g=0^+$ to the attractive regime $\g=0^{-}$~\cite{Koch2021}: once the bound states are formed, they remain stable due to approximate integrability as interactions become more attractive, while their binding energy grows. Eventually, in a realistic experiment~\cite{horvath2025}, highly-energetic bound states become unstable and cause atom losses. Roughly, this can be expected to happen when the binding energy of the bound states becomes comparable with the transverse trapping energy, and the 1D regime is lost.
The GHD equations in the presence of bound states are analogous to Eqs. \eqref{Seq_GHD_rho} \eqref{Seq_GHD_fill}, where an extra index should be added labeling the bound states $\rho(\lambda)\to \{\rho_j(\lambda)\}_{j=1}^N$. Here, each $\rho_j(\lambda)$ represents the root density associated with a bound state (bright solitons) of $j$ particles. By retaining only $j=1$, the GHD equations \eqref{Seq_GHD_rho} \eqref{Seq_GHD_fill} are obtained.
A complete discussion of these equations can be found in Ref.~\cite{Koch2022}, as well as the generalization to Eq.~\eqref{Seq_Etot} and \eqref{Seq_g2} plotted in Fig.~4.
In Fig.~\ref{Fig_S1}, we show the bound state population $P_j\equiv \int \dd\lambda\, \rho_j(\lambda)$ obtained by reverse crossing the non-interacting point after entering the $\mathcal{W}=1,2,3$ cycles according to the analytical results from Ref.~\cite{Koch2021}. Notice that, crossing $\g=0$ from repulsive to attractive, the entropy is continuous~\cite{Koch2021}.

\section{Sampling the momentum distribution via Monte Carlo method}
\label{sec_MC}

This section gives an overview of the numerical method used to compute the momentum distribution and ultimately $g^{(1)}(x)$ through the Fourier transform.
The method proposed in Ref.~\cite{senese2025} combines known analytical formulas from integrability~\cite{Calabrese2016} with a Monte Carlo approach.
With minor modifications, we follow Ref.~\cite{senese2025} and provide a short overview of the method for completeness.
We aim to compute $\langle \hat{\psi}^\dagger(x) \hat{\psi}(0)\rangle$ on a GGE. Using the microscopic states \eqref{eq_suppstate} this amounts to

\begin{multline}\label{Seq_g1}
\langle \hat{\psi}^\dagger(x) \hat{\psi}(0)\rangle=\frac{1}{\mathcal{Z}_\text{GGE}}\sum_{\{\lambda_j\}_{j=1}^N}\langle \{\lambda_j\}_{j=1}^N| \hat{\psi}^\dagger(x) \hat{\psi}(0)|\{\lambda_j\}_{j=1}^N\rangle e^{-\sum_{j=1}^n \left(\sum_k \beta_k q_k(\lambda_j)\right)}=\\
\frac{1}{\mathcal{Z}_\text{GGE}}\sum_{\{\lambda_j\}_{j=1}^N}\sum_{\{\mu_\ell\}_{\ell=1}^{N-1}} |\langle \{\mu_j\}_{\ell=1}^{N-1}|\hat{\psi}(0)|\{\lambda_j\}_{j=1}^N\rangle |^2e^{i x \hbar^{-1} \left(\sum_{j=1}^{N} p(\lambda_j)-\sum_{\ell=1}^{N-1} p(\mu_\ell)\right)}e^{-\sum_{j=1}^n \left(\sum_k \beta_k q_k(\lambda_j)\right)}
\end{multline}

Analytical expressions as determinants of $(N-1)\times (N-1)$ matrices are available for the form factor $|\langle \{\mu_j\}_{\ell=1}^{N-1}|\hat{\psi}(0)|\{\lambda_j\}_{j=1}^N\rangle |^2$~\cite{Calabrese2016}: for completeness, we also report them
at the end of this section.
Importantly, these determinant representations allow an efficient computation of the form factor for a large number of particles, eventually ensuring the feasibility of the Monte Carlo approach. 
It is now convenient to collect the form factor and GGE weight in an effective Boltzmann weight
\be
\mathcal{E}_\text{MC}(\{\mu_j\}_{\ell=1}^{N-1}|\{\lambda_j\}_{j=1}^{N})=-\log|\langle \{\mu_j\}_{\ell=1}^{N-1}|\hat{\psi}(0)|\{\lambda_j\}_{j=1}^N\rangle |^2 +\sum_{j=1}^n \left(\sum_k \beta_k q_k(\lambda_j)\right)\, .
\ee
Notice that, as previously discussed, by reversing Eq.~\ref{eq_tba}, starting with a known form of $\rho(\lambda)$ describing a GGE, we can recover the weight function $\sum_k \beta_k q_k(\lambda)$. This is indeed the procedure we use to determine the GGE's weight $\sum_k \beta_k q_k(\lambda)$ during the cycle.
From a direct inspection of Eq.~\eqref{Seq_g1} notice that $n=\langle \hat{\psi}^\dagger(0)\hat{\psi}(0)\rangle=\mathcal{Z}_\text{GGE}^{-1} \mathcal{Z}_\text{MC}$, where we define $\mathcal{Z}_\text{MC}=\sum_{\{\lambda_j\}_{j=1}^N}\sum_{\{\mu_\ell\}_{\ell=1}^{N-1}}e^{-\mathcal{E}_\text{MC}(\{\mu_j\}_{\ell=1}^{N-1}|\{\lambda_j\}_{j=1}^{N})}$.
With this observation, we conveniently move to the Fourier space in defining the momentum distribution as $P(p)=\int \frac{\dd x}{2\pi}\, e^{-i x\hbar^{-1} p}\langle \hat{\psi}^\dagger(x) \hat{\psi}(0)\rangle$, which can now be represented as

\be\label{eq_Pdis}
\frac{1}{n}P(p)=
\frac{1}{\mathcal{Z}_\text{MC}}\sum_{\{\lambda_j\}_{j=1}^N}\sum_{\{\mu_\ell\}_{\ell=1}^{N-1}} \delta\left(p\hbar^{-1}- \hbar^{-1} \left(\sum_{j=1}^{N} p(\lambda_j)-\sum_{\ell=1}^{N-1} p(\mu_\ell)\right)\right)e^{-\mathcal{E}_\text{MC}(\{\mu_j\}_{\ell=1}^{N-1}|\{\lambda_j\}_{j=1}^{N})}
\ee

We can now use a Markov chain to sample the above right hand side, updating the configurations $(\{\mu_j\}_{\ell=1}^{N-1}|\{\lambda_j\}_{j=1}^{N})$ accordingly as follows.

\begin{enumerate}
\item We first update the $\{\lambda_j\}_{j=1}^N$ configuration, parametrizing through the quantum numbers $\{I_j\}_{j=1}^N$ numerically solving the Bethe equations \eqref{Seq_bethe}. 
A single step consists in: 
\begin{enumerate}
\item Randomly select a label $\bar{j}$ and update $I_j\to I'_{\bar{j}} =I_j+\delta I_{\bar{j}}$. $\delta I_{\bar{j}}$ is chosen as a random integer in the interval $[-\delta_\text{max},\delta_\text{max}]$, the value of $\delta_\text{max}$ is tuned to retain at least $10\%$ acceptance rate.
\item First one checks if the new configuration is allowed, ensuring $I'_{\bar{j}}\ne I_j$ for every $j$. If the new configuration is allowed, the new rapidity configuration $\{\lambda'_j\}$ is computed solving the Bethe Equations. The new configuration is accepted with probability $\exp\left[- \sum_k \beta_k q_k(\lambda'_j)+\sum_k \beta_k q_k(\lambda_j)\right]$. Notice that computing this acceptance probability is fast as it does not require computing the form factor, and the computational cost is mainly due to solving the Bethe equations.
\item Repeat the above procedure for a number of $n_\text{GGE}$ of proposed allowed configurations.
\end{enumerate}
\item After a sufficient number of updates of the $\lambda-$configurations, we update the $\mu-$configurations.
\begin{enumerate}
\item First a seed is proposed. One randomly selects a label $\bar{j}$ and defines a set of quantum numbers $\{J_\ell\}_{\ell=1}^{N-1}$ parametrizing the $\mu-$rapidities by removing from the $\{I_j\}_{j=1}^N$ set the quantum number $I_{\bar{j}}$, and then the others are shifted of $-1/2$ if they precede $I_{\bar{j}}$, and of $+1/2$ if they follow, to accommodate for the changed parity in the number of particles of the state with a hole. The probability used to choose the hole position $\bar{j}$ is determined by the weights $e^{-\mathcal{E}_\text{MC}(\{\mu_j\}_{\ell=1}^{N-1}|\{\lambda_j\}_{j=1}^{N})}$.
\item After the hole is created, the initial configuration is updated $\{\mu_\ell\}_{\ell=1}^{N-1}\to \{\mu'_\ell\}_{\ell=1}^{N-1}$ similarly to what previously done with the $\lambda-$rapidities, but determining the acceptance probability with $\exp\left[-\mathcal{E}_\text{MC}(\{\mu'_j\}_{\ell=1}^{N-1}|\{\lambda_j\}_{j=1}^{N})+\mathcal{E}_\text{MC}(\{\mu_j\}_{\ell=1}^{N-1}|\{\lambda_j\}_{j=1}^{N})\right]$.
\item The $\mu-$rapidities are updated for $n_\text{FF}$ allowed proposed moves. For each $\lambda-$configuration, the momentum distribution \eqref{eq_Pdis} is sampled only after a $10\%$ of the initial $n_\text{FF}$ steps have been performed, to avoid the initial bias.
Fig.~\ref{Fig_S_conv} shows, for a fixed configuration of $\{\lambda_i\}$, the typical convergence in the evolving $\{\mu_i\}_{i=1}^{N-1}$.
\begin{figure}[h!]
\includegraphics[width=0.99\columnwidth]{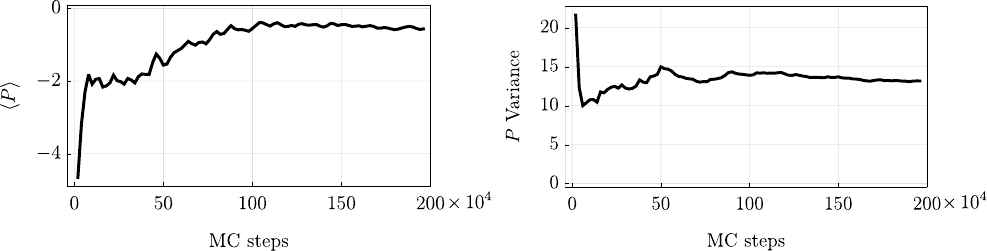}
\caption{\textbf{Convergence of the Monte Carlo sampling in the number of steps.---} For the example of $\gamma=5$ and $N=30$ number of particles, shows the average of the momentum $P$ (left) and its variance (right) through the Markov chain evolving the $\{\mu_j\}_{j=1}^{N-1}$ configuration.
}\label{Fig_S_conv}
\end{figure}
\end{enumerate}
\item Step 1 and 2 are repeated $n_\text{sample}-$times, averaging each time.
\item Errors are estimated by running independent copies of the Markov chains, considering their average as the most representative value.
\item For the data shown in this work, we consider $N=30$ particles and $L=30$, in such a way the density is one $n=1$. We checked convergence to the thermodynamic limit, within the Monte Carlo fluctuations, upon varying $N$ and $L$. A comparison of $g^{(1)}$ obtained with $N=30$ and $N=50$ is provided in Fig. \ref{Fig_S_conv_size}, showing convergence in $N$ is attained within the uncertainty of the Monte Carlo sampling.

For each fixed $\{\lambda_i\}_{i=1}^N$ we consider $2\times 10^6$ updates of the $\{\mu_i\}_{i=1}^{N-1}$ configuration which is sampled every $2\times 10^4$ steps. The $\{\lambda_i\}_{i=1}^N$ configuration is sampled every $5\times 10^5$ steps. With a choice of $\delta_\text{max}=4$, the acceptance is around $10\%$, but it changes for different $\gamma$. For each point of the cycle parametrized by $(\gamma,\wind)$, we run $5$ independent realization of the Monte Carlo, each made of about $1000$ samples. To show statistical uncertainty, these partial averages are shown in the plots with light shading together with their average (solid colors), which is the most representative value.
\end{enumerate}

\begin{figure}[t!]
\includegraphics[width=0.75\columnwidth]{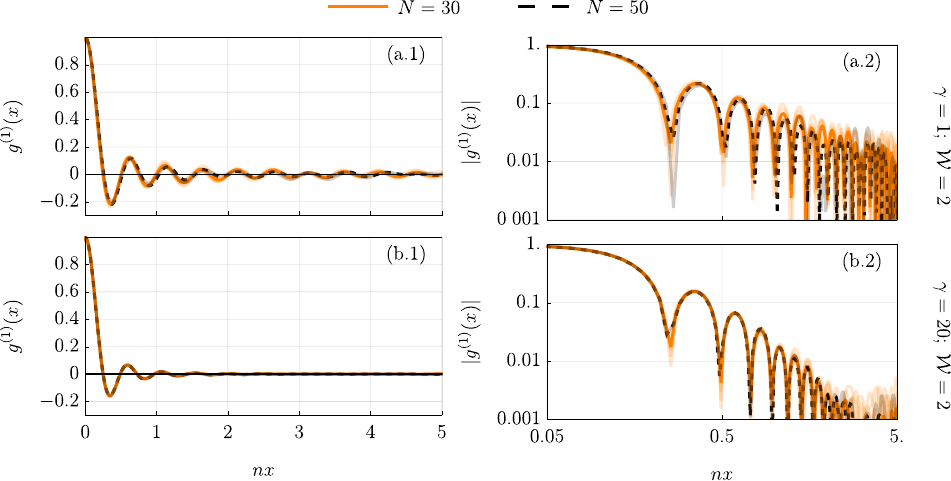}
\caption{\textbf{Convergence of the Monte Carlo sampling in the number of particles---} For the examples of $(\gamma=1,\wind=2)$ ((a.1) and (a.2)) and $(\gamma=20,\wind= 2)$ ((b.1) and (b.2)), we show the convergence of the Monte Carlo sampling upon varying the number of particles $N$ and the size $L$, fixing the density $n=N/L=1$. We show results for $N=30$ (solid orange line) and $N=50$ (dashed black line). For a fair comparison, we consider the same number of Monte Carlo samplings: we consider 1000 independent samplings, divided in 5 batches of 200 to assess statistical fluctuations. Light orange and black lines, more evident in the log scale (a.2) and (b.2), are the partial averages for $N=30$ and $N=50$, respectively. Other choices of $\gamma$ and $\wind$ show a similar overlap for the two values of $N$, proving convergence. Except for this figure, the $N=30$ case is always considered with a larger total number of samples (5000), to further reduce statistical uncertainity.
}\label{Fig_S_conv_size}
\end{figure}
\begin{figure}[b!]
\includegraphics[width=0.99\columnwidth]{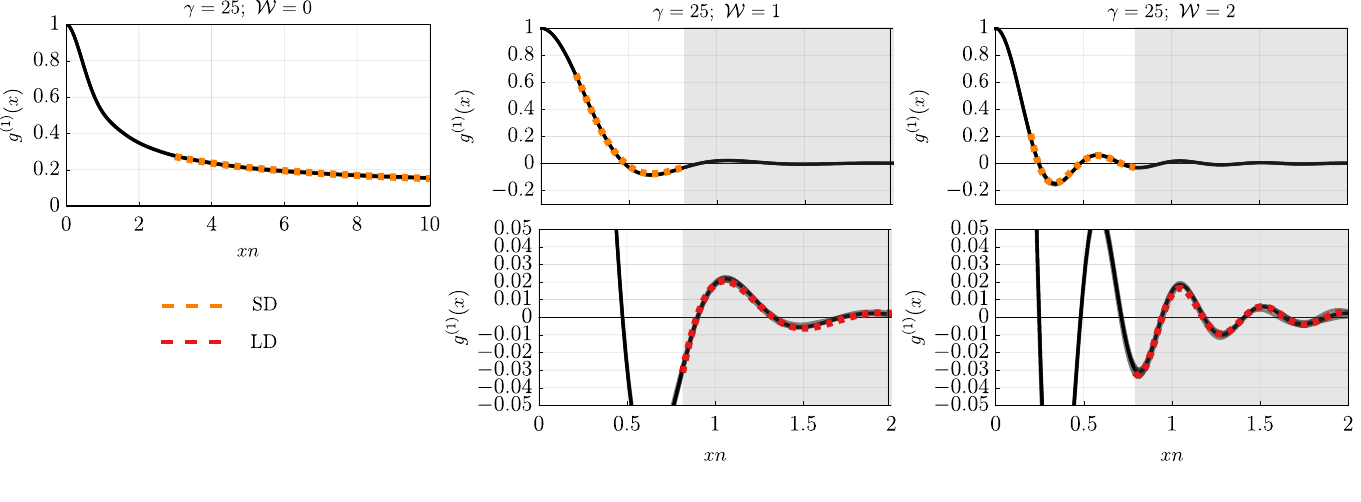}
\caption{\textbf{Fitting the oscillating decay in $g^{(1)}$---} We show a typical example of the fit of the bimodal oscillating power-law displayed by $g^{(1)}(x)$. Numerical data from Monte Carlo sampling are shown in black for $30$ particles and $n=1$. To assess the statistical error,
we consider five independent Monte Carlo samplings (gray lines) of 1000 samples each, and show their average as the solid black curve.
We focus on $\gamma=25$ and different branches of the cycle $\wind=0,1,2$ (from left to right). In the upper row we show the fit (yellow line) of the short distance (SD) power law, and highlight the interval used for the fit (shaded area). In the second row, we zoom-in on the large distance (LD) tails which are independently fitted (red line). For $\wind=0$ the tails are not shown, since there is no change in the observed power law.} \label{Fig_S_Osc}
\end{figure}

After having sampled the momentum distribution, $g^{(1)}(x)$ is recovered through Fourier transform. Then, we fit the oscillating power law decay: we now further describe our fitting procedure.
For $\wind=0$, where FO are subdominant, we fit $g^{(1)}(x)$ with a simple power law $A /x^B$. Instead, for the oscillating $g^{(1)}(x)$ featured for $\wind\ge 1$, we use $A \cos(k_\text{FO} x+\phi)/x^\nu$.
The fit is performed at short distances within an interval $[x_{1,a},x_{2,a}]$ and then at larger distances $[x_{1,b},x_{2,b}]$ to capture the short-distance and large-distance power laws. In practice, for $\wind\ge 1$ we choose $[x_{1,a},x_{2,a}]=[0.05,\bar{x}]$ and $[x_{1,b},x_{2,b}]=[\bar{x},\tilde{x}]$. 
We discuss below how $\bar{x}$ and $\tilde{x}$ are determined.
An example of the fit is given in Fig.~\ref{Fig_S_Osc} for the complete fitting function, while in the main text we focused only on the power law.
The crossover distance between the two-power laws $\bar{x}$ is defined by the intersection of the power-law envelope. After having fitted the short-distance (SD) and long-distance (lD) oscillating powerlaws with the convention that $A_\text{SD}>0$ and $A_\text{LD}>0$, we define
\be
\bar{x}=\exp\left[\frac{\log A_\text{SD}-\log A_\text{LD}}{\nu_\text{LD}-\nu_\text{SD}}\right]\, .
\ee
The large cutoff $\tilde{x}$ is instead determined as it follows.
We observe that for values of $|g^{(1)}(x)|\lesssim 10^{-3}$ the statistical error becomes of the same order of the correlation itself , hence we determine $\tilde{x}$ from the threshold such that the power-law envelope becomes smaller than the cutoff $A_\text{LD}/\tilde{x}^{\nu_\text{LD}}=10^{-3}$.
The values of $\bar{x}$ and $\tilde{x}$ are self-consistently determined: we repeat the fit several times adjusting the fitting interval until convergence is reached.

\begin{figure}[h!]
\includegraphics[width=0.99\columnwidth]{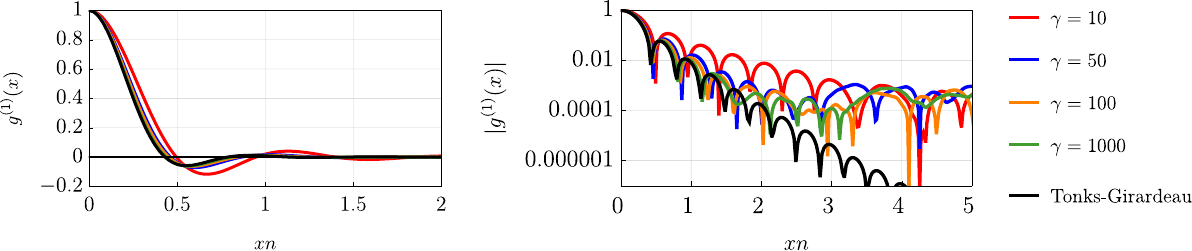}
\caption{\textbf{Tonks-Girardeau regime and transition to exponential decay---} For the example $\wind=1$, the $g^{(1)}(x)$ for four strongly interacting datasets from the Monte Carlo sampling are shown, respectively for $\gamma=10, 50,100$ and $1000$ (for simplicity, only the average is shown without error bars) and compared against the Tonks-Girardeau limit. In the linear scale (left) the Monte Carlo data converge to the Tonks-Girardeau limit for strong repulsion. In the right panel we show the data in a log scale, emphasizing the Tonks-Girardeau limit clearly features an oscillating exponential decay. As commented in the main text, the apparent saturation of the Monte Carlo data to a plateau is a numerical artifact of the limited statistical precision of the method. 
}\label{Fig_S_TG}
\end{figure}

We observe that as $\gamma$ is increased, the power decay becomes faster until at strong $\gamma$ is difficult to clearly distinguish the two power-laws. In the Tonks-Girardeau regime, we observe a transition between the bimodal power law to an oscillating exponential decay. In the Tonks-Girardeau regime, $g^{(1)}(x)$ can be alternatively computed taking advantage of the fermionization of the hard-core bosons. This allows to compute $g^{(1)}(x)$ with high precision --despite the quickly-vanishing tails-- directly in the thermodynamic limit without need of the Monte Carlo routine described above, see eg. Ref.~\cite{horvath2025}.
In Fig.~\ref{Fig_S_TG}.
In Fig.~\ref{Fig_S_TG}, we present Monte Carlo results at large interaction strength together with the Tonks–Girardeau limit of infinite interactions, that shows a clear exponential decay.

\subsection{The Form Factor}
\label{SSec_FF}

The determinant form for the form factor $|\langle \{\mu_j\}_{\ell=1}^{N-1}|\hat{\psi}(0)|\{\lambda_j\}_{j=1}^N\rangle |^2$ has been computed in Ref.~\cite{Caux2007} as (below, $c\equiv m \g/\hbar^2)$
\be\label{eq_FFA}
|\langle \{\mu_j\}_{\ell=1}^{N-1}|\hat{\psi}(0)|\{\lambda_j\}_{j=1}^N\rangle |^2= c^{2N-1}\frac{\prod_{j>k=1}^N((\lambda_j-\lambda_k)^2+c^2)^2}{\prod_{a=1}^N \prod_{b=1}^{N-1}(\lambda_a-\mu_b)^2}\frac{\det U(\{\mu_i\}_{i=1}^{N-1},\{\lambda_j\}_{j=1}^N)}{||\{\lambda_j\}_{j=1}^N||^2||\{\mu_i\}_{i=1}^{N-1}||^2}\, ,
\ee
where $||\{\lambda_j\}_{j=1}^N||^2$ is the norm of a Bethe state that is computable as
\be
||\{\lambda_j\}_{j=1}^N||^2=c^N\prod_{j>k=1}^N\frac{(\lambda_j-\lambda_k)^2+c^2}{(\lambda_j-\lambda_k)^2} \det \mathcal{G}(\{\lambda_j\}_{j=1}^N)\hspace{2pc}\text{with} \hspace{2pc}
\mathcal{G}(\{\lambda_j\}_{j=1}^N)=\delta_{j,k}\left[L+\sum_{\ell=1}^N\varphi(\lambda_j-\lambda_\ell)\right]-\varphi(\lambda_j-\lambda_k)\, .
\ee
The matrix $U(\{\mu_i\}_{i=1}^{N-1},\{\lambda_j\}_{j=1}^N)$ appearing in Eq.~\eqref{eq_FFA} is a $N-1\times N-1$ matrix with elements
\be
 U_{j,k}(\{\mu_i\}_{i=1}^{N-1},\{\lambda_j\}_{j=1}^N)=\delta_{j,k}\left(\frac{V_j^+-V_j^-}{i}\right)+\frac{\partial_{a=1}^{N-1}(\mu_a-\lambda_j)}{\partial_{a\ne j}^{N-1}(\lambda_a-\lambda_j)} \big[\varphi(\lambda_j-\lambda_k)-\varphi(\lambda_N-\lambda_k)\big]\, ,
\ee
where $V_j^\pm=\prod_{a=1}^{N-1}(\mu_a-\lambda_j \pm i c)/\prod_{a=1}^N(\lambda_a-\lambda_j\pm i c)$.

\end{document}